\documentclass{aa}  

\usepackage{graphicx}
\usepackage{txfonts}
\usepackage{comment}
\usepackage[colorlinks=true,allcolors=blue]{hyperref}
\usepackage{xcolor}
\usepackage{enumitem}

\newcommand{\chandra}{\textit{Chandra}}
\newcommand{\xmm}{\textit{XMM-Newton}}

\newcommand{\swift}{\textit{Swift}}
\newcommand{\nustar}{\textit{NuSTAR}}

\newcommand{\ergcms}{erg cm$^{-2}$ s$^{-1}$}
\newcommand{\ergs}{erg s$^{-1}$}

\begin{document}

   \title{The ultraluminous X-ray source M81 X-6: a weakly magnetised neutron star with a precessing accretion disc?}
    \titlerunning{The ULX M81 X-6: a weakly magnetised NS with a precessing accretion disc?}
   \subtitle{}

   \author{R. Amato
          \inst{1}
          \and A. G\'urpide
          \inst{1,2}
          \and N. A. Webb
          \inst{1}
          \and O. Godet
          \inst{1}
          \and M. J. Middleton
          \inst{2}
          }

   \institute{IRAP, CNRS, Université de Toulouse, CNES, 9 Avenue du Colonel Roche, 31028 Toulouse, France \and  School of Physics \& Astronomy, University of Southampton, Southampton, Southampton SO17 1BJ, UK \\
              \email{roberta.amato@irap.omp.eu} }

   \date{}

\abstract{Thanks to their proximity, ultraluminous X-ray sources (ULXs) represent a privileged astrophysical laboratory to study super-Eddington accretion. Current open questions concern the nature of the compact object, which is still hard to determine in those cases where pulsations are not directly detected, and the mechanisms responsible for the spectral changes observed in many ULXs.}
{We investigate the nature of the ULX M81 X-6, which has been suggested to harbour a neutron star (NS), by studying its long-term X-ray spectral and temporal evolution, with the goal of assessing the astrophysical phenomena responsible for its spectral changes.}
{Using the rich set of available archival data from \xmm, \chandra, \nustar,\ and \swift/XRT, we tracked the evolution of the source on the hardness-intensity diagram and inferred the different emitting regions of the system and their geometry, as well as the mechanisms responsible for the spectral transitions.}
{We find that the source oscillates between two main states: one characterised by a hard and luminous spectrum and the other at low hardness and luminosity. The properties of the soft component remain constant between the two states, suggesting that changes in the mass-transfer rate are not driving the spectral transitions. Instead, the bi-modal behaviour of the source and the known super-orbital period would point to the precession of the accretion disc. Here, we tested two theoretical models: (1) Lense-Thirring precession, which can explain the super-orbital period if the NS has a magnetic field $B$ $\lesssim10^{10}$ G, supporting the idea of M81 X-6 as a weakly magnetised NS, and (2) precession due to the torque of the NS magnetic field, which leads to $B \gtrsim$ 10$^{11}$ G. However, the latter scenario, assuming M81 X-6 shares similar properties with other NS-ULXs, is disfavoured because it would require magnetic field strengths ($B>10^{15}$ G) much higher than those known for other pulsating ULXs. We further show that the contribution from the hard component attributed to the putative accretion column sits just below the typical values found in pulsating ULXs, which, together with the low value of the pulsed fraction ($\leq10$\%) found for one \xmm/pn observation, could explain the source's lack of pulsations.}
{The spectral properties and variability of M81 X-6 can be accounted for if the accretor is a NS with a low magnetic field. Under the hypothesis of Lense-Thirring precession, we predict a spin period of the NS of a few seconds. We encourage future X-ray pointed observations to look for pulsations and/or spectral signatures of the magnetic field.} 

 

   \keywords{Accretion, accretion discs -- Stars: neutron -- X-rays: binaries -- X-rays: individuals: M81 X-6}
             
   \maketitle

%

\section{Introduction}
\label{sec:intro}

Ultraluminous X-ray sources (ULXs) -- defined as extragalactic, off-nuclear, point-like X-ray sources with luminosities in excess of the Eddington limit for a stellar-mass black hole (BH; $L_X\gtrsim10^{39}$ \ergs, under the assumption of isotropic emission) --  have been explained as X-ray binaries powered by accretion at super-Eddington mass-transfer rates \citep[see][for a review]{Kaaret2017}. They have long been thought to host intermediate-mass BHs \citep[see e.g.][]{Mezcua2017}, but the discovery of coherent pulsations from one of them \citep[M82 X–2;][]{Bachetti2014} showed that ULXs can also be powered by accretion onto stellar-mass compact objects. Since then, five more pulsating ultraluminous X-ray sources  \citep[PULXs;][]{Fuerst2016,Israel2017a,Israel2017b,Carpano2018,RodriguezCastillo2020,Sathyaprakash2019} and a few candidates \citep[e.g.][]{Quintin2021} have been identified. At present, the ULX zoo is diverse, with the most luminous ones, called hyperluminous X-ray sources \citep[][$L_X>10^{41}$ \ergs]{Gao2003}, still consistent with accreting intermediate-mass BHs \citep[e.g. ESO 243-49 HLX-1,][]{Farrell2009} and those with luminosities in the range $10^{39}-10^{41}$ \ergs\ consistent with accreting stellar-mass BHs or neutron stars (NSs), with the NS-ULX population possibly dominant \citep[e.g.][]{KingLasota2016,Koliopanos2017,Middleton2017,Pintore2017,Walton2018}.

The spectral characteristics of ULXs suggest that observations may be influenced by two major factors: the mass-accretion rate \citep[][]{Sutton2013} and the inclination of the system with respect to the observer, with a possible precession of the accretion disc \citep[][]{Middleton2015}. According to the most accepted picture, matter accreting onto the compact object forms a disc with a complex structure: it starts as a standard thin accretion disc \citep{ShakuraSunyaev1973} in the outermost part and becomes geometrically thick, inflated by the radiation pressure, until reaching its maximum height at the so-called spherisation radius, where the local radiation flux is equal to the Eddington flux. From the spherisation radius inwards, powerful outflows, which become clumpy at a certain distance, are launched from the surface of the disc \citep{ShakuraSunyaev1973,Poutanen2007, Takeuchi2013}.  Following this idea, ULX X-ray spectra have been modelled with phenomenological components, with the aim to assess the different parts of the accretion disc. One possibility involves two multi-temperature disc models \citep[][G21 hereafter]{Walton2018,Walton2018b,Gurpide2021a}, one with temperatures $\gtrsim$1 keV, usually attributed to the innermost part of the accretion disc, and one with lower temperatures, attributed to the outermost regions of the disc \citep[but see also][for a different interpretation of the components]{Koliopanos2017}. Blueshifted absorption lines with velocities of a fraction of the speed of light at soft energies seen in a few ULXs \citep{Middleton2014, Middleton2015b, Walton2016b,Pinto2016, Pinto2020, Pinto2021,Kosec2018,Kosec2021} would be the signature of the powerful ionised gas outflows. For now, the detection of these lines is at the limit of what current X-ray instrumentation is capable of, with detections at only the few sigma level.

On the basis of their spectral shape, ULXs are usually classified into three main categories: hard ultraluminous (HUL), soft ultraluminous (SUL), and broadened disc \citep[BD;][]{Sutton2013}. The HUL and SUL sources can be modelled with a double black body, with temperatures around 0.1--0.3 keV and 1--3 keV for the cool and hot component, respectively. The dominance of one component with respect to the other determines the nature of HUL (hot component) or SUL (cool component) sources. On the other hand, the BD-ULX spectra can be modelled with a disc model with a radial temperature profile $T(R)\propto R^{-p}$, with $p<0.75$, indicative of an advection-dominated disc \citep{Mineshige1994}, as opposed to a standard thin disc, which would have $p=0.75$ \citep{ShakuraSunyaev1973}. In addition, a sub-class of super-soft ultraluminous sources (SSUL) has also been identified, with ULXs exhibiting spectra dominated almost entirely by soft emission below $\sim2$ keV \citep[see e.g.][]{Urquhart&Soria2016}. Moreover, while some sources seem to permanently stay in the same class, others transit from one to another \citep{Sutton2013,Middleton2015,
Gurpide2021b}. 

M81 X-6 is a bright ULX ($L_\mathrm{X}\gtrsim 
2\times10^{39}$ \ergs) located in the outskirts of its host galaxy. Discovered first by \cite{Fabbiano1988}, this source was immediately classified as a ULX. The first observation of the source with the \textit{Chandra} X-ray observatory \citep{Weisskopf2000} was analysed by \cite{Swartz+2003}, who fitted the spectrum with a disc black-body model of $\sim1$ keV. Later, \cite{Gladstone2009} noticed a characteristic curvature above 2 keV and made the analogy of the spectral shape being similar to a BH binary in the hard state. \citet{Sutton2013} modelled the spectrum as a BD, but noticed that the source was still consistent with the HUL regime, especially considering its X-ray luminosity of $\gtrsim3\times10^{39}$ \ergs. Recently, \citet{Jitesh2018} and \citet{Jithesh2020} conducted a broadband study of the source, analysing X-ray data from \xmm{} \citep{Jansen2001}, \nustar{} \citep{Harrison2013}, and \textit{Suzaku} \citep{Mitsuda2007}. Their broadband spectral analysis revealed similarities between M81 X-6 and other PULXs, but a search for periodic modulation failed to identify any strong signal. They concluded that the source might be a potential PULX candidate.
Finally, G21 explored this source in the context of a more general study on the long-term X-ray spectral evolution of ULXs. By tracking the evolution of the source in the hardness-intensity diagram (HID), they showed how the source goes back and forth between two main states. By comparing the evolution of the hardness with those of known PULXs, the authors identified M81 X-6 as the best NS-ULX candidate of their sample. A similar bi-modal behaviour was also observed by \cite{Weng&Feng2018} using observational data from the X-ray Telescope \citep[XRT;][]{Burrows2005} on board the \textit{Neil Gehrels Swift} Observatory \citep{Gehrels2004}. 

All of the aforementioned spectroscopic studies have used X-ray data taken up to 2012 and/or two simultaneous \nustar{} and \textit{Suzaku} observations taken in 2015. Here, we aim to explore the evolution of M81 X-6 with a more thorough and up-to-date analysis that accounts for all the available \chandra\ and \xmm\ archival data from 2012 to 2020, complemented by all the data from \swift/XRT from its launch up to June 2021. In particular, this study benefits from several unpublished \chandra\ and \xmm\ observations carried out in two different years (2016-2017 for \chandra\ and 2020 for \xmm; see Table\,\ref{tab:LogObs}), which guarantee, for the first time, a spectral monitoring of the source over weekly, monthly, and yearly timescales. 
We also used the aforementioned \nustar{} dataset to investigate the spectrum at high energies (see Sect. \ref{sec:spectral_analysis}).

\begin{table*}
      \caption[]{Log of \chandra\ and \xmm\ observations considered in this work. \xmm\ exposure times have been cleaned by high-background intervals.}
         \label{tab:LogObs}
     $$ 
         \begin{tabular}{llcccc}
         \hline\hline\noalign{\smallskip}
            Observatory & ObsID & Date & Instr. & Exp. (ks) & Net counts \\
            \noalign{\smallskip}
            \hline\hline
            \noalign{\smallskip}
            \chandra & 18052 & 2016-02-01 & ACIS-I & 20.2 & 1492 \\
            &18054 & 2016-06-21 & ACIS-I & 65.71 & 5825 \\
            &18875 & 2016-06-24 & ACIS-I & 32.36 & 4236 \\
            &18047 & 2016-07-04 & ACIS-I & 100.43 & 12570 \\
            &18817 &	2016-08-07 & ACIS-I & 39.55 & 2978 \\
            &19685 &	2016-08-08 & ACIS-I & 36.59 & 2264 \\
            &18053 & 2017-01-05 & ACIS-I & 29.59 & 3296 \\
            &19982 & 2017-01-12 & ACIS-I & 67.49 & 8510 \\
            &18051 &	2017-01-24 & ACIS-I & 36.58 & 6504 \\
            &19993 &	2017-01-26 & ACIS-I & 24.75 & 4116 \\
            &19992 &	2017-01-28 & ACIS-I & 34.61 & 5967 \\
            \noalign{\smallskip}
            \hline
            \noalign{\smallskip}
            \xmm & 0843840101 & 2020-03-18 & MOS2 & 22.72 & 2078\\
            &0870930101 & 2020-10-17 & MOS1/MOS2 & 22.27/22.28 & 1012/781 \\
            &0870930301 & 2020-10-29 & MOS2 & 15.65 & 503 \\
            &0870930401 & 2020-11-06 & MOS1/MOS2 & 5.43/4.9 & 159/118\\
            &0870900101 & 2020-11-14 & pn/MOS2 & 17.15/24.45 & 4087/1815 \\
            &0870900201 & 2020-11-16 & pn/MOS1/MOS2 & 12.74/25.25/25.24 & 3854/2423/2544 \\
            &0870930501 & 2020-11-18 & MOS1 & 23.6 & 913 \\
            &0870931001 & 2020-11-24 & MOS1/MOS2 & 24.5/20.35 & 829/640 \\
            \noalign{\smallskip}
            \hline\hline
         \end{tabular}
     $$ 
   \end{table*}

\section{Data reduction}

\subsection{\chandra}

M81 was observed by \chandra\ more than 15 times between February 2016 and January 2017. However, the source was within the field of view in only 11 observations, reported in Table \ref{tab:LogObs}. All observations were reprocessed with the \textit{Chandra} Interactive Analysis of Observations \citep[CIAO; version 4.13,][]{Fruscione2006} software package, using the script \texttt{chandra\char`_repro}, with  calibration files CALDB version 4.9.5. For each observation, we extracted spectra and light curves, using circular extraction regions of 15$^{\prime\prime}$ for the source and of 45$^{\prime\prime}$ for the background, respectively. The unusually large radius for the source region was due to either the intensity of the source or to its off-axis position on the ACIS CCDs. We made sure that both the source and background extraction regions were on the same CCD to account for a correct subtraction of the background. Spectra were binned with a minimum of 20 counts/bin and light curves were extracted in three different energy bands: broad (0.3--10 keV), hard (1.5--10 keV), and soft (0.3--1.5 keV). Light curves were barycentred at the source coordinates RA=9:55:32.9 and Dec=$+$69:00:33  \citep[J2000,][]{Gladstone2009} We also checked for pile-up, using the CIAO tool \texttt{pileup\char`_map}\footnote{\url{https://cxc.harvard.edu/ciao/ahelp/pileup_map.html}.}, finding a level of pile-up $<$5\% in all the observations.

\subsection{\xmm}
\label{sub:xmmdata}
\xmm\ archival data up to 2012 were already analysed in depth by G21. Therefore, we focused on more recent observations. \xmm\ observed the galaxy M81 14 times in 2020, but the ULX fell into the field of view in only eight observations (see Table\,\ref{tab:LogObs}). Six observations have no pn data because they were not centred on M81 X-6 and the pn was in the small window or timing mode. \xmm\ data were reduced with the Science Analysis System (SAS; version 19.1.0) and the respective current calibration files, and reprocessed with the standard SAS routines \texttt{emproc/epproc}, filtering the 10--12 keV light curve for count rates $>$0.4 counts s$^{-1}$ to prevent contamination from the high-flaring background. Response files were generated using the tasks \texttt{rmfgen} and \texttt{arfgen}.
We extracted spectra and barycentred light curves as described above for the \chandra\ data. Unfortunately, the \xmm{} observations are much shorter than the \chandra\ observations, preventing us from using them for the timing analysis (cf. Sect.\,\ref{sec:timing_analysis}).

\subsection{\swift/XRT}

\begin{figure*}
    \resizebox{\hsize}{!}
    {\includegraphics[]{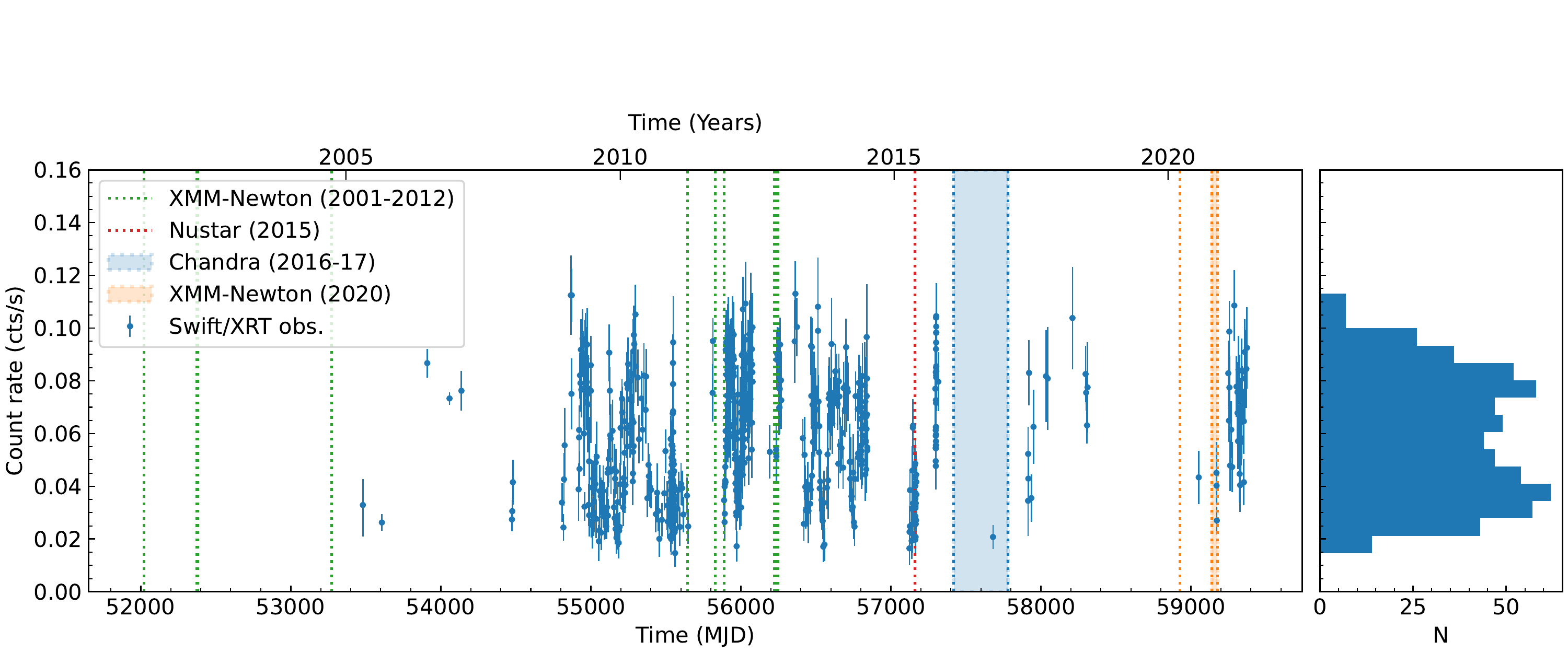}}
    \caption{Temporal evolution of M81 X-6. \textit{Left panel}: \swift/XRT light curve in the 0.3--10 keV energy band. Each data point corresponds to an observation. The time interval spun by the \chandra\ observations considered in this work is coloured in light blue, while that of the most recent \xmm\ observations is coloured in light orange or represented by a dotted orange vertical line. The vertical dotted green lines mark the oldest \xmm\ observations (cf.\,G21) and the red one the \nustar\ observation used in this work. \textit{Right panel}: Histogram of \swift/XRT count rates, showing a clear bimodal distribution.}
    \label{fig:Swiftlc}
\end{figure*}

With the goal of exploring the gap between the latest \chandra\ and \xmm\ observations and those in G21, we looked into the \swift\ data archive\footnote{\url{https://www.swift.ac.uk/swift_live/index.php}.}. For each of the 603 observations, we used the online tools\footnote{\url{https://www.swift.ac.uk/user_objects/}.} \citep{Evans2007,Evans2009} to build the light curves in the same broad, hard, and soft energy bands to retrieve the hardness of the source. Figure\,\ref{fig:Swiftlc} shows the broadband (0.3--10. keV) \swift/XRT light curve, with the times of \chandra, \xmm, and \nustar\ observations highlighted.

\subsection{\nustar}
\nustar\ has observed the galaxy M81 in seven instances, catching the ULX only in one of the observations (ObsID 60101049002). The data were extracted with the \nustar\ Data Analysis Software version 2.0.0 with \textit{CALDB} version 20200826. Source and background spectra were extracted using the task \textit{nuproducts} with the standard filters. Source events were selected from a circular region centred on the source of 60" in radius, to avoid contamination from the nearby active galactic nucleus in M81. Background regions were selected from larger, circular, source-free regions and on the same chip as the source but as far away as possible to avoid contamination from the source itself. \nustar\ spectra were re-binned to 40 counts per bin due to its lower energy resolution and the lower signal-to-noise ratio compared to the EPIC cameras.

\section{Spectral analysis}
\label{sec:spectral_analysis}

As mentioned in Sect. \ref{sec:intro}, M81 X-6 shows spectral changes between states at high and low luminosity. To better understand and eventually classify the source spectral states, as a first step, we decided to follow the method of G21 and fit each individual \chandra\ and \xmm\ spectrum with two double-absorbed disc black bodies. All fits were performed with the X-ray spectral fitting package Xspec \citep{Arnaud1996}, version 12.12.0b, with abundances from \citet{Wilms2000} and cross-sections from \citet{Verner1996}. The two black-body components were reproduced by the Xspec model \texttt{diskbb} \citep{Makishima1986,Mitsuda1984}, so that the overall fit function was: \texttt{tbabs*tbabs*(diskbb+diskbb)}. The Galactic absorption was fixed to $9.8\times10^{20}$ cm$^{-2}$ \citep{HI4PICollaboration2016} and local absorption was fixed to $11.7\times10^{20}$ cm$^{-2}$, the average value determined by G21, since it was not always possible to constrain it. \xmm{} observations where data from more than one instrument (MOS1, MOS2 and/or pn) were present simultaneously were fitted together. We did not perform a spectral fit only for the \xmm{} ObsID 0870930401, due to the low number of counts. Best-fit values are reported in Tables\,\ref{tab:BestFitChandra} and \ref{tab:BestFitXMM}, with errors at 90\% confidence level.

As already observed in G21 for M81 X-6, as well as for several other ULXs, the fit with two thermal black-body components is in general a good fit, with reduced $\chi^2$ values close to 1, except for a few observations, where it is lower than 1 (noticeably \chandra\ ObsIDs 18052 and 19993, and \xmm{} ObsIDs 0870930101, 0870930501, and 0870931001, most likely because of the low number of counts). Unfortunately, because of the quality of the data, for several \chandra\ observations the best-fit values were not always constrained and in many cases we were only able to place lower limits. Overall, the best-fit temperatures of the soft component are around 0.4 keV, but while they are well constrained for XMM-Newton, those of Chandra have very large uncertainties.
The best-fit values of the hard component are always higher than 1.3 keV. Both best-fit temperatures are in general consistent with those found by G21 for the 2001--2012 \xmm\ datasets, as well as the unabsorbed luminosities in the range $\sim2-5\times10^{39}$ \ergs{}.

We noticed that in almost all \chandra\ observations and in two \xmm\ ones (ObsID 0870900101 and 0870900201) residuals at energies $\lesssim1$ keV are marginally evident. Those residuals have sometimes been interpreted as signatures of powerful outflows irradiated from the surface of the accretion disc \citep{Middleton2014, Middleton2015b}. In order to resolve the spectroscopic features of these winds, high-resolution spectra are necessary \citep{Pinto2016,Pinto2020,Pinto2021,Kosec2021}. Hence, we did not add any additional feature to the fit model.

\begin{table*}
      \caption[]{Best-fit parameters for each \chandra\ observation.} 
         \label{tab:BestFitChandra}
     $$ 
     \begin{small}
         \begin{tabular}{lccccccccc}
            \hline\hline\noalign{\smallskip}
            Obs.ID & $kT_\mathrm{soft}$ (keV) & Norm. & $kT_\mathrm{hard}$ (keV) & Norm. ($10^{-2}$) & Flux (10$^{-12}$ \ergcms) & Lum. (10$^{39}$ \ergs)& $\chi^2_\mathrm{red}$/dof & State\\
            \noalign{\smallskip}
            \hline\hline
            \noalign{\smallskip}
            G21 & 0.4--0.6 & 0.3--2 & 1.4--2.5 & 0.1--3 & & 2.4--7.4 & 0.6--1.2\\
            \hline\noalign{\smallskip}
            18052 & $0.60\pm0.15$ & $0.28_{-0.14}^{+0.43}$ & $1.7$\tablefootmark{b} & $0.07_{-3.4}^{+1.3}$ & $1.44\pm0.06$ & $2.2\pm0.1$  & 0.72/78 & SLL\\
            \noalign{\smallskip}
            18054 & $0.73\pm0.09$ & $0.19_{-0.05}^{+0.08}$ & $3$\tablefootmark{b} & $0.02_{-0.02}^{+0.09}$ & $2.14\pm0.05$ & $3.34\pm0.07$ & 0.90/206 & SLL\\
            \noalign{\smallskip}
            18875 & $0.8\pm0.3$ & $0.14_{-0.07}^{+0.36}$ & $1.6$\tablefootmark{b} & $0.4_{-0.4}^{+1.2}$ & $2.89\pm0.07$ & $4.15\pm0.11$ & 0.93/182 & HHL \\
            \noalign{\smallskip}
            18047 & $0.35_{-0.08}^{+0.12}$ & $2.28_{-1.6}^{+5.6}$ & $1.65_{-0.09}^{+0.14}$ & $1.8\pm0.5$ & $3.26\pm0.05$  & $5.08\pm0.08$ & 0.98/321 & HHL\\
            \noalign{\smallskip}
            18817 & $0.46_{-0.12}^{+0.15}$ & $0.8_{-0.5}^{+1.7}$ & $2.0_{-0.5}^{+1.8}$ & $0.3_{-0.3}^{+0.6}$ & $1.69\pm0.05$ & $2.64\pm0.08$ & 0.93/140 & SLL\\
            \noalign{\smallskip}
            19685 & $0.40_{-0.07}^{+0.09}$ & $1.5_{-0.9}^{+2.1}$ & $1.8_{-0.3}^{+0.5}$ & $0.5_{-0.3}^{+0.5}$ & $1.66\pm0.05$ & $ 2.59\pm0.09$ & 1.01/129 & SLL\\
            \noalign{\smallskip}
            18053 & $0.4_{-0.3}^{+5.1}$ & $0.7_{-0.6}^{+12}$ & $1.3$\tablefootmark{b} & $3.4_{-3.4}^{+1.2}$ & $2.80\pm0.08$ & $4.36\pm0.13$ & 0.98/154 & HHL\\
            \noalign{\smallskip}
            19982 & $0.32_{-0.17}^{+0.56}$ & $1.18_{-1.11}^{+20}$ & $1.45_{-0.07}^{+0.27}$ & $3.3_{-2.0}^{+0.7}$ & $3.17\pm0.06$ & $4.94\pm0.09$ & 0.97/274 & HHL\\
            \noalign{\smallskip}
            18051 &	$0.6_{-0.6}^{+3.3}$ & 0.1\tablefootmark{a} & $1.4_{-0.1}^{+4.0}$ & $3.4_{-3.4}^{+1.3}$ & $3.25\pm0.07$ & $5.1\pm0.1$ & 1.13/232 & HHL\\
            \noalign{\smallskip}
            19993 & $1.17$\tablefootmark{a} & $0.06$\tablefootmark{a} & $2.8$\tablefootmark{a} & $0.075$\tablefootmark{a} & $3.08\pm0.08$ & $4.81\pm0.13$ & 0.84/180 & HHL\\
            \noalign{\smallskip}
            19992 &	$0.74$\tablefootmark{a} & $0.06$\tablefootmark{a} & $1.4$\tablefootmark{b} & $2.4$\tablefootmark{a} & $3.04\pm0.07$ & $4.74\pm0.11$ & 1.01/216 & HHL\\
            \noalign{\smallskip}            \hline\hline
         \end{tabular}
        \end{small}
        $$ 
     \tablefoot{The interstellar absorption is fixed to $9.8\times10^{20}$ cm$^{-2}$ and the local one to $11.7\times10^{20}$ cm$^{-2}$ (G21). Errors are at 90\%. Fluxes and luminosities  are both unabsorbed and computed in the 0.3--10 keV range. The range of the best-fit values from G21 are reported in the first row for an easy reference to the reader. The last column reports whether the source was caught in the hard/high-luminosity (HHL) or soft/low-luminosity (SLL) regime.
     \tablefoottext{a}{Parameter not constrained.}
     \tablefoottext{b}{Lower limit.}
     }
   \end{table*}

\begin{table*}
      \caption[]{Best-fit parameters for each \xmm{} observation.}
         \label{tab:BestFitXMM}
     $$ 
     \begin{small}
         \begin{tabular}{lccccccccc}
            \hline\hline\noalign{\smallskip}
            Obs.ID & $kT_\mathrm{soft}$ (keV) & Norm. & $kT_\mathrm{hard}$ (keV) & Norm. ($10^{-2}$) & Flux (10$^{-12}$ \ergcms) & Lum. (10$^{39}$ \ergs)& $\chi^2_\mathrm{red}$/dof & State\\
            \noalign{\smallskip}
            \hline\hline
            \noalign{\smallskip}
            G21 & 0.4--0.6 & 0.3--2 & 1.4--2.5 & 0.1--3 & & 2.4--7.4 & 0.6--1.2\\
            \noalign{\smallskip}
            \hline
            \noalign{\smallskip}
            0843840101 & $0.43_{-0.12}^{+0.16}$ & $0.8_{-0.5}^{+1.6}$ & $1.7_{-0.3}^{+0.4}$ & $1.4_{-0.8}^{+1.2}$ & $2.95\pm0.11$ & $4.59\pm0.17$ & 0.97/85 & HHL \\
            \noalign{\smallskip}
            0870930101 & $0.43_{-0.07}^{+0.08}$ & $1.1_{-0.5}^{+1.0}$ & $2.4_{-0.7}^{+2.4}$ & $0.17_{-0.16}^{+0.5}$ & $1.79\pm0.07$ & $2.79\pm0.11$ & 0.86/79 & SLL \\
            \noalign{\smallskip}
            0870930301 & $0.33_{-0.09}^{+0.12}$ & $2.2_{-1.5}^{+5.3}$ & $2.3_{-0.8}^{+3.5}$ & $0.2_{-0.2}^{+0.7}$ & $1.58\pm0.13$ & $2.5\pm0.2$ & 1.11/23  & SLL\\
            \noalign{\smallskip}
            0870930401 & -- & -- & -- & -- & -- & -- & -- & SSL \\
            \noalign{\smallskip}
            0870900101 & $0.45_{-0.05}^{+0.06}$ & $0.9_{-0.3}^{+0.5}$ & $1.9_{-0.2}^{+0.4}$ & $0.6_{-0.3}^{+0.5}$ & $2.30\pm0.05$ & $3.59\pm0.08$ & 1.15/150 & SLL \\
            \noalign{\smallskip}
            0870900201 & $0.42\pm0.03$ & $1.2_{-0.3}^{+0.4}$ & $2.0_{-0.2}^{+0.3}$ & $0.32_{-0.14}^{+0.21}$ & $1.64\pm0.03$ & $2.57\pm0.05$ & 1.15/233 & SLL \\
            \noalign{\smallskip}
            0870930501 & $0.43\pm0.11$ & $1.0_{-0.6}^{+1.7}$ & $2.4_{-0.9}^{+6.9}$ & $0.17_{-0.17}^{+0.9}$ & $1.6\pm0.1$ & $2.57\pm0.15$ & 0.77/45 & SLL \\
            \noalign{\smallskip}
            0870931001 & $0.40_{-0.05}^{+0.06}$ & $1.4_{-0.6}^{+1.1}$ & $4.5_{-2.1}^{+4.5}$ & $0.02_{-0.02}^{+0.14}$ & $1.75\pm0.08$ & $2.73\pm0.13$ & 0.84/78 & SLL \\
            \noalign{\smallskip}            \hline\hline
         \end{tabular}
        \end{small}
        $$ 
   \end{table*}

\begin{figure*}
    \resizebox{\hsize}{!}
    {\includegraphics[]{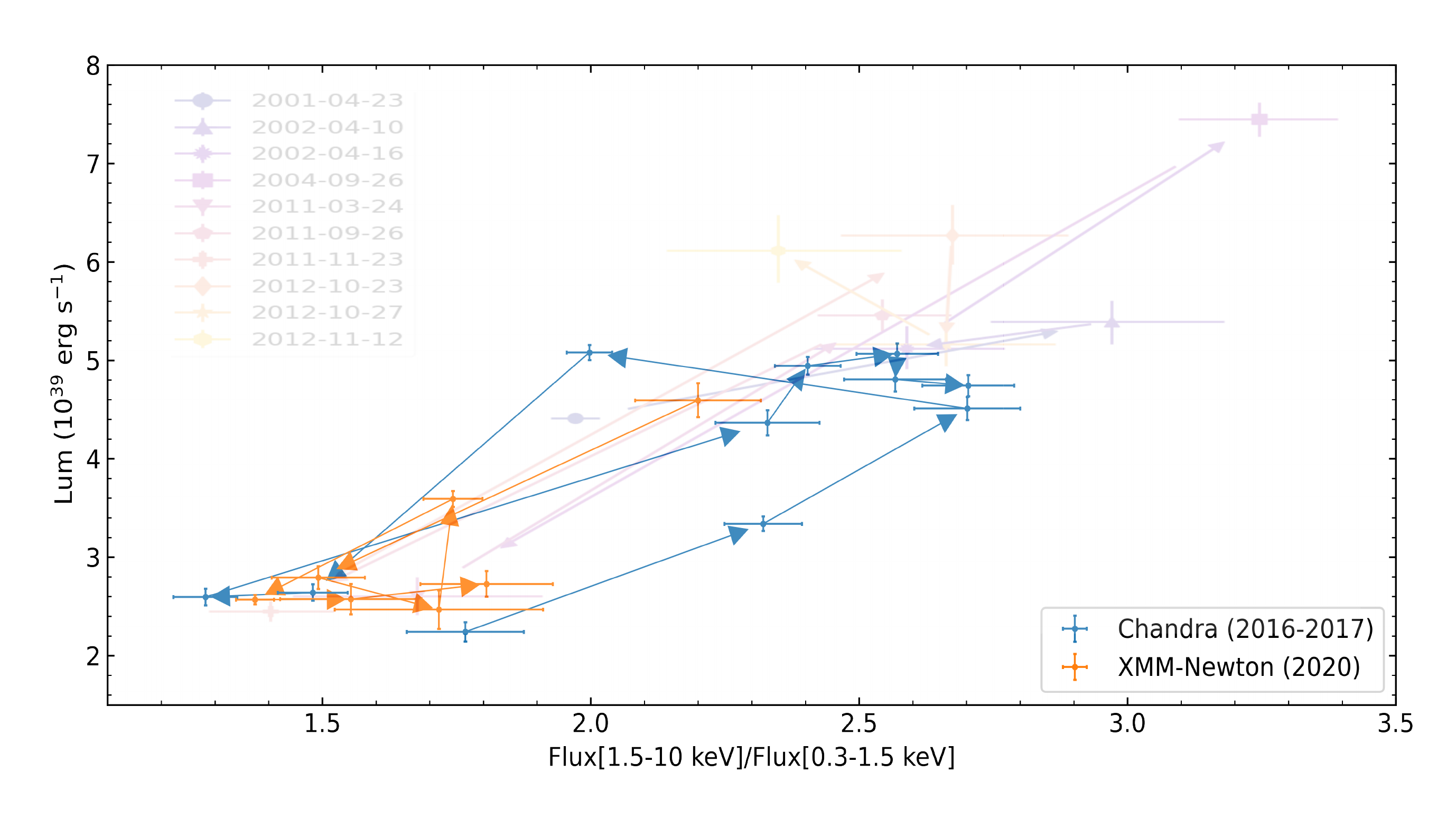}}
    \caption{Hardness-intensity diagram (HID) of the \chandra\ (blue) and 2020 \xmm\ (orange) datasets, in comparison with the old \xmm\ observations (fainter text and points) as reported in G21.}
    \label{fig:LumFluxXMM}
\end{figure*}

As in G21, we also computed the hardness ratio (HR) as the ratio between the unabsorbed flux in the hard band (1.5--10 keV) over the unabsorbed flux in the soft band (0.3--1.5 keV) and plotted it in the HID of Fig.\,\ref{fig:LumFluxXMM}, where we also show the results from G21 for an easy comparison. Our results are consistent with those of G21, except for a single old \xmm\ observation, taken in 2004, where the source was caught with a higher luminosity ($\sim8\times10^{39}$ \ergs), while in the recent observations it never went above $L_X\sim5\times10^{39}$ \ergs.

\begin{figure*}
    \resizebox{\hsize}{!}
    {\includegraphics[]{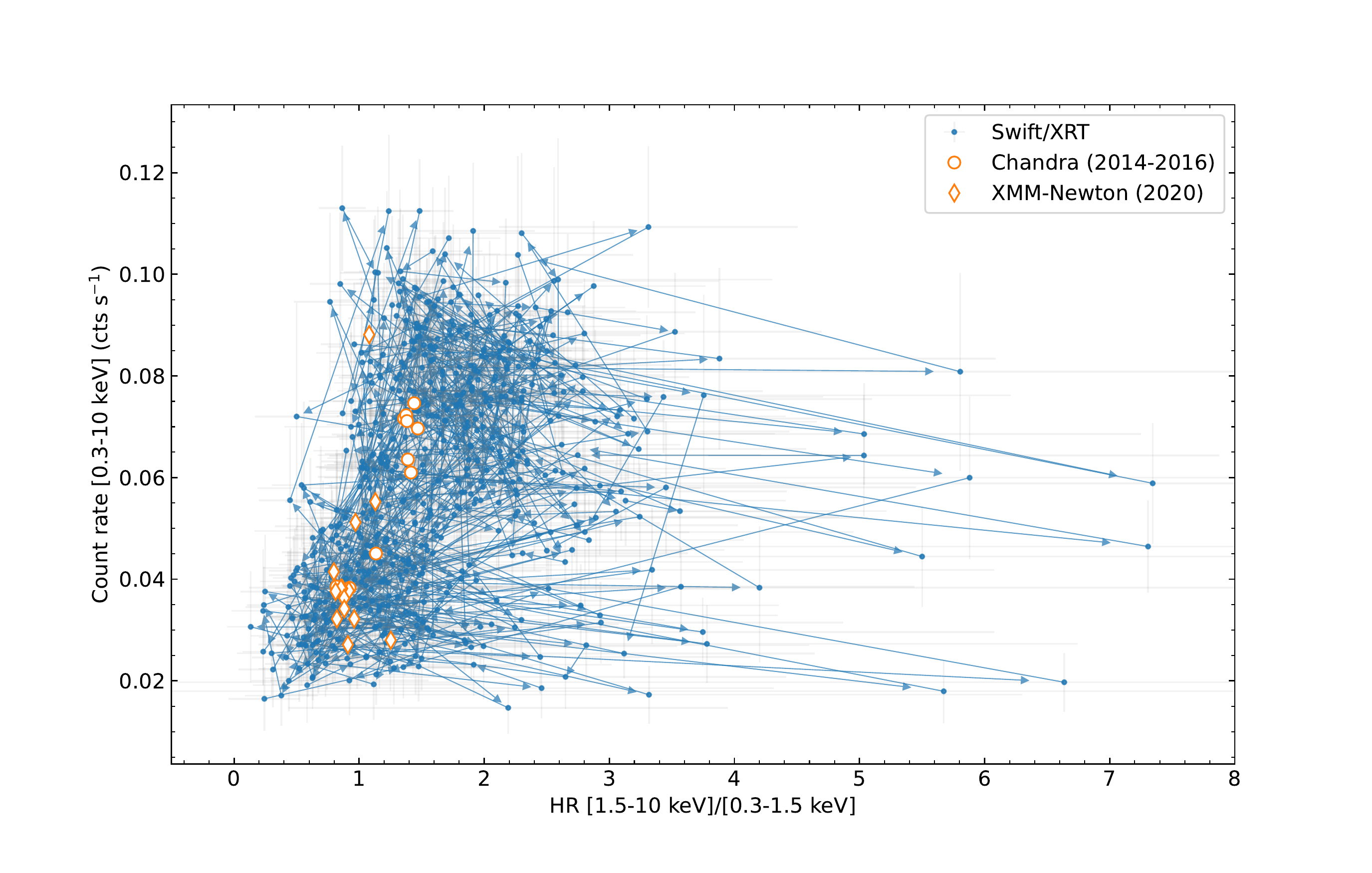}}
    \caption{Evolution of \swift/XRT broadband count rates (blue dots) as a function of the HR, in comparison with the latest \chandra\ (empty orange circles) and \xmm\ data (empty orange diamonds). Data from different \xmm-EPIC instruments are plotted individually. Errors on \chandra\ and \xmm\ data are not shown.}
    \label{fig:Swifthardness}
\end{figure*}

We also investigated whether M81 X-6 entered other regions of the HID, as seen in Holmberg II X–1 and NGC 5204 X–1 \citep{Gurpide2021b}. To this end, we considered all \swift/XRT archival data up to June 2021. Following \citet{Gurpide2021b}, we computed the HR for each observation as the ratio of the counts in the hard (1.5--10 keV) and soft (0.3--1.5 keV) energy bands. To directly compare \swift/XRT data with those from the other two X-ray satellites, we converted \chandra\ and \xmm\ count rates into \swift/XRT count rates. Briefly, we convolved the absorbed best-fit models of each \chandra/\xmm\ observation with the \swift/XRT redistribution matrix (swxpc0to12s6\_20130101v014.rmf) and ancillary file
(swxs6\_20010101v001.arf) of 01/08/2020. Results are shown in Fig.\,\ref{fig:Swifthardness}, where the broadband (0.3--10 keV) count rates are plotted against the HR. \chandra\ and \xmm\ data are perfectly consistent with \swift/XRT with no overall evidence of a third state, where the source is supposed to be harder and/or at higher luminosities. Our analysis confirms the results of \cite{Weng&Feng2018}.

\begin{figure}
    \resizebox{\hsize}{!}
    {\includegraphics[]{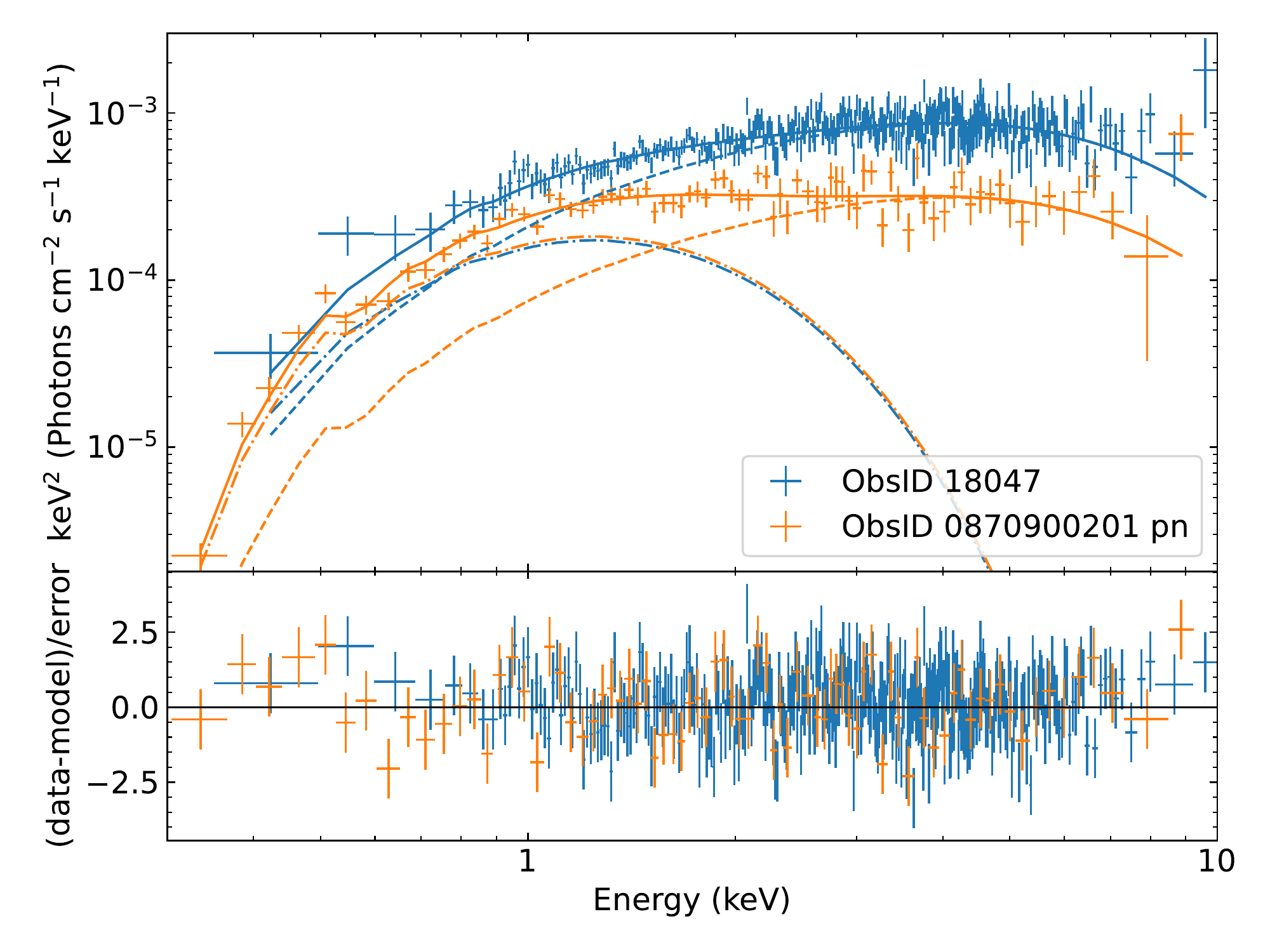}}
    \caption{Spectra of \chandra\ ObsID 18047 (blue) and \xmm/pn ObsID 0870900201 (orange), representative of the HHL and SLL states, respectively. Spectra are fitted with the model \texttt{tbabs*tbabs*(diskbb+diskbb)} (solid line) with the two black-body components plotted as dashed lines. \textit{Bottom panel:} Residuals of the fit. }
    \label{fig:SampleSpectra}
\end{figure} 

From the inspection of the individual \chandra\ and \xmm\ spectra and of the HIDs, we found that the source transits between two main regimes: a hard, high-luminosity (HHL) state, characterised by an HR$\sim$2--3 and luminosities of the order of 4--6$\times10^{39}$ \ergs{}, and a soft, low-luminosity (SLL) state, with HR$\lesssim$2 and $L_\mathrm{X}\sim3\times10^{39}$ \ergs{} (see Tables \ref{tab:BestFitChandra} and \ref{tab:BestFitXMM}). 
Figure\,\ref{fig:SampleSpectra} shows two spectra representative of each group. We note that these two regimes do not fit in the traditional HUL and SUL states (cf. Sect.\,\ref{sec:intro}). The overall hardness of the source is always $\gtrsim1$ (see e.g. Figs.\,\ref{fig:LumFluxXMM}, \ref{fig:Swifthardness}, and \ref{fig:18047hr}), which would classify M81 X-6 as a HUL source. We simply note that, it is within the same HUL category that M81 X-6 explores two different regimes. These two regimes are also evident in the double-peaked distribution of \swift/XRT count rates (Fig.\,\ref{fig:Swiftlc}, right panel).

Once we had established the boundaries of the two spectral states seen for M81 X-6, we grouped together the spectra of the same type and fitted them, with the aim to constrain the temperatures and the normalisations of the two black bodies. At this stage, to obtain statistically acceptable fits, we had to exclude \chandra\ ObsIDs 18875 (HHL) and 18054 (SLL) and \xmm\ ObsIDs 0870900101 and 0870930401 (SLL),  because they represented transitory stages between the two regimes. Best-fit results of the state-resolved spectra are reported in Table \ref{tab:BestFitGroups}. The best-fit temperatures resulted in $T_\mathrm{hard}=1.46_{-0.05}^{+0.17}$ keV and $T_\mathrm{soft}=0.45_{-0.15}^{+0.51}$ keV for the HHL regime, and $T_\mathrm{hard}=2.2_{-0.2}^{+0.3}$ keV and $T_\mathrm{soft}=0.48_{-0.04}^{+0.05}$ keV for the SLL regime. These results would suggest that the soft component does not vary between the two regimes, in agreement with G21. Thanks to the high number of total counts, this time we were able to constrain also the local absorption, which we left free to vary. It resulted in $n_H=(11.2_{-5.0}^{+8.1})\times10^{20}$ cm$^{-2}$ and $n_H=(7.7_{-2.1}^{+2.3})\times10^{20}$ cm$^{-2}$, for the HHL and the SLL regimes, respectively, consistent in both cases with the average value of $11.7\times10^{20}$ cm$^{-2}$ estimated by G21.

Whilst these last fits are statistically acceptable, we noticed that the best-fit temperature of the hard component is higher in the SLL state rather than in the HHL state. This may imply that the hard part of the spectrum is not well-described by the dual thermal component model, as suggested by weak residuals at energies $\gtrsim$8 keV seen in a few spectra (e.g. ObsIDs 18054, 18047, 18817, 19685). In the literature, an additional cut-off power law has often been used by several authors to model high-energy excesses and it has been linked to the X-ray radiation from the accretion column of the NS in PULXs \citep[e.g.][cf. Sect. \ref{sec:discussion}]{Walton2018}. Hence, we decided to add this further cut-off power law component (\texttt{cutoffpl} in Xspec) to the best-fit model, thus allowing for the possibility of a NS accretor also for M81 X-6. To constrain the parameters of the cut-off power law, we used the \nustar\ observation, taken when M81 X-6 was in the SLL regime (see Fig.\,\ref{fig:Swiftlc}). Hence, at first, we fitted all the \xmm\ observation in the SLL state used in the previous step together with the \nustar/FPMA and FPMB data, with an absorbed double disc black-body plus cut-off power law model (i.e. \texttt{tbabs*tbabs*(diskbb+diskbb+cutoffpl)}). \nustar\ data were limited to the 3--20 keV range, because of the high background contamination at higher energies. To compute the errors on $E_\mathrm{cut}$ and $\Gamma$, we froze the local absorption and alternately one of the two parameters, leaving the other free to vary. We found the best-fit values of the high-energy cut-off to be $E_\mathrm{cut}=4.3_{-0.3}^{+1.3}$ keV and for the power law index $\Gamma=0.38_{-0.90}^{+0.15}$. These values are consistent with those of other ULXs \citep[see e.g.][]{Walton2018}. To assess whether adding a cut-off power law actually improves the fit, we performed an F-test, which resulted in a statistical value of 4.2 and a probability of 0.04, meaning that the additional cut-off power law component is marginally significant. Next, we froze the power law parameters to their best-fit values and fitted all the spectra in the two regimes with the same model, with the local absorptions, the two black-body temperatures and all the normalisations tied for each group. Best-fit results are shown in Table \ref{tab:BestFitGroups}. We also performed an F-test for the HHL dataset, which was inconclusive. This is not surprising, since the cut-off power law is well characterised for the SLL state but not for HHL, due to the lack of high-energy coverage of this latter state. Nonetheless, we chose to keep the cut-off power law also for the HHL state, in light of the interpretation of it modelling the X-ray emission coming from the accretion column of the NS. As a matter of fact, because we expect the cut-off power law to be stronger in the HHL state, the best-fit values found for the SLL state could be considered lower limits for the parameters of the cut-off power law in the HHL state. In any case, whether we take the cut-off power law into account or not, the best-fit values of the black-body components of the HHL state remain unchanged. Overall, we found the best-fit temperatures of the soft black-body component of the two regimes in agreement with each other ($T_\mathrm{soft}\approx0.4$ keV), while the best-fit temperature of the hard black body seems to be higher in the HHL state ($T_\mathrm{hard}=1.46_{-0.15}^{+0.16}$ keV) and lower in the SLL state ($T_\mathrm{hard}=0.9_{-0.2}^{+0.8}$ keV), though they are consistent within the uncertainties.

To further constrain the soft component, we performed one last, simultaneous fit of the two groups, where we linked the temperature and the normalisation of the soft component for all the observations. We left free to vary within each group the local absorption, the temperature and normalisation of the hard component, and the normalisation of the cut-off power law. We obtained a fit with $\chi^2_\mathrm{red}$(dof)$=1.06(1985)$ and a null hypothesis probability (n.h.p.) of 0.03. The best-fit temperature of the soft component was $T_\mathrm{soft}=0.34_{-0.05}^{+0.06}$ keV, with a normalisation of $1.4_{-0.8}^{+1.7}$. As expected, this best-fit temperature is consistent with the best-fit temperatures of the soft component of the two states fitted separately. All the other best-fit parameters were also consistent with the previous fits. Although the value of the n.h.p. would imply that this fit is only marginally statistically acceptable, the consistency of all the best-fit parameters with the previous fits reassures on its validity. Nevertheless, in the successive steps of our analysis we consistently use the best-fit of $T_\mathrm{soft}\sim0.4$ keV, as obtained in the separate fits of the two regimes.

\begin{figure*}
    \resizebox{\hsize}{!}
    {\includegraphics[]{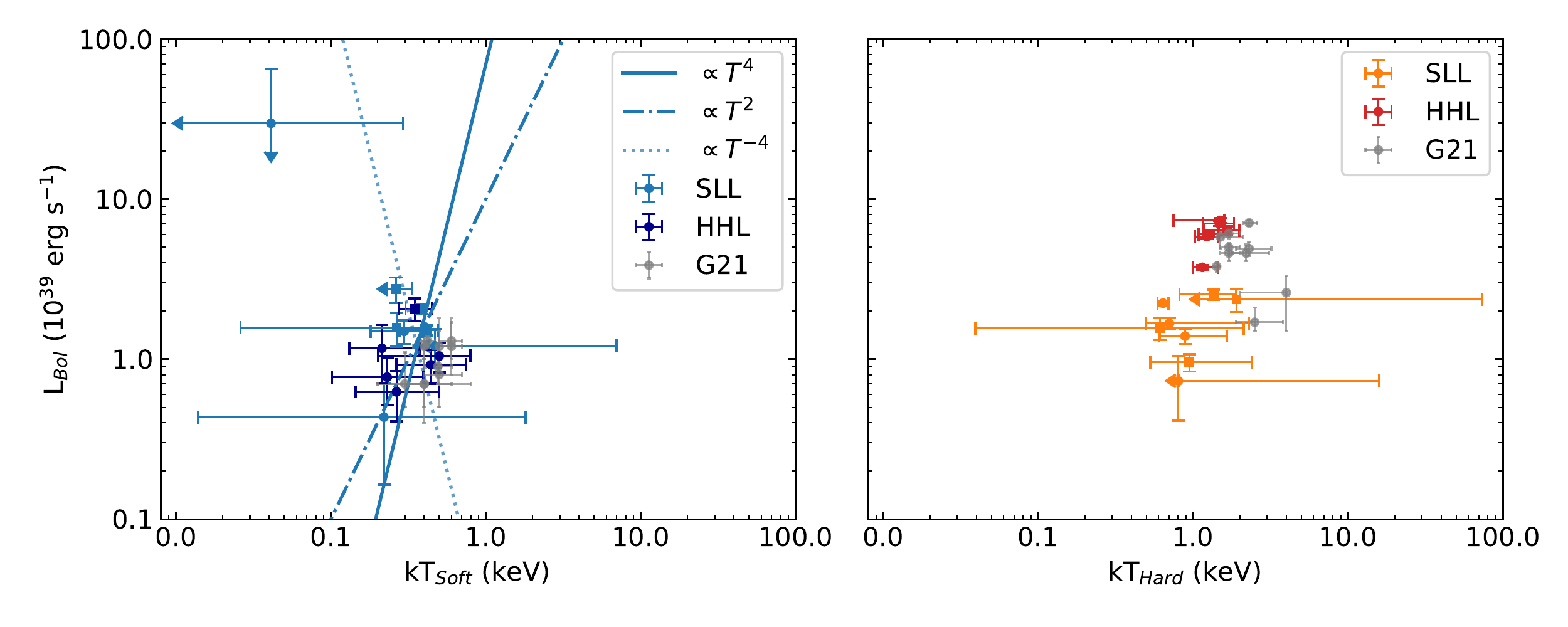}}
    \caption{Bolometric luminosity as a function of the temperature for the soft (left panel) and hard (right panel) black-body components, in comparison with the older \xmm\ observations as in G21 (in grey). For the soft component, three $L-T$ laws are overplotted: $L\propto T^4$ for a thin disc \citep{ShakuraSunyaev1973}, $L\propto T^2$ for an advection-dominated disc around a BH \citep{Watarai2000}, and $L\propto T^{-4}$ as observed for NGC 1313 X-2 \citep{Robba2021}. The bolometric luminosities are computed in the 0.01--100 keV range.}
    \label{fig:LumBolTemp}
\end{figure*}

We also studied the relationship of the hard and soft temperatures with their unabsorbed bolometric luminosities (in the range 0.01--100 keV). Results are shown in Fig.\,\ref{fig:LumBolTemp}, together with values from G21.  
We did not find any trend, neither for the hard nor the soft component. Nonetheless, we plot the expected $L-T$ relations of $L\propto T^4$ in the case of a thin disc \citep{ShakuraSunyaev1973} and $L\propto T^2$ for an advection-dominated disc around a BH \citep[see e.g.][]{Watarai2000,Walton2020}, as well as the trend $L\propto T^{-4}$ recently proposed for NGC 1313 X-2 \citep{Robba2021}. The lack of any trend for our dataset is not surprising, as we are attempting to use simple models to fit ULX spectra when the physics involved is certainly more complex. Moreover, as remarked in G21, 
these correlations are likely due to the degeneracy between the temperature and the normalisation of each component, casting doubts on their physical interpretations.

As mentioned in Sect. \ref{sec:intro}, the HHL state of M81 X-6 has sometimes been also modelled with a BD component \citep[e.g.][]{Sutton2013}, such as \texttt{diskpbb} in Xspec \citep{Mineshige1994}. To test this model, we tried to substitute the harder multi-temperature disc component with a BD for the spectra in HHL state, so that the overall model is now: \texttt{tbabs*tbabs*(diskbb+diskpbb+cutoffpl)}. We note that the \texttt{diskpbb} alone can likely reproduce the entire 0.3--10 keV spectrum, but, because we wanted to test a model with two thermal components, we kept them both and froze the parameters of the soft component to ensure its significant contribution to the fit. In this way, we obtained a best-fit temperature of the BD component of $kT=1.50_{-0.19}^{+0.07}$ keV, a parameter $p=0.70_{-0.03}^{+0.07}$, and a normalisation of $2.4_{-0.6}^{+2.0}\times10^{-2}$, with a $\chi^2_\mathrm{red}$(dof)$=1.06(1151)$ and a n.h.p. of 0.08. The parameter $p$ is consistent at 90\% confidence level with the value of 0.75 for a thin Shakura-Sunyaev disc. Hence, there is only marginal evidence of the hard disc component to be advection-dominated. We found similar results for the SLL spectra. For this reason, we do not investigate further and consider the model with two disc black bodies plus the cut-off power law as the best-fit model.

\begin{table*}
      \caption[]{Best-fit parameters of the observations in the two regimes.}
         \label{tab:BestFitGroups}
     $$ 
         \begin{tabular}{lcccc}
            \hline\hline\noalign{\smallskip}
            & HHL & SLL & HHL & SLL \\
            \noalign{\smallskip}
            &&& \texttt{+cutoffpl} & \texttt{+cutoffpl} \\
            \noalign{\smallskip}
            \hline
            \noalign{\smallskip}
            $N_H$ ($10^{20}$ cm$^{-2}$) & $11.2_{-5.0}^{+8.1}$ & $7.7_{-2.1}^{+2.3}$ & $11.5_{-5.0}^{+7.9}$ & $9.40\pm0.03$ \\
            \noalign{\smallskip}
            $kT_\mathrm{soft}$ (keV) & $0.45_{-0.15}^{+0.51}$ & $0.48_{-0.04}^{+0.05}$ & $0.4_{-0.2}^{+0.5}$ & $0.39\pm0.09$ \\
            \noalign{\smallskip}
            Norm. & $0.3_{-0.2}^{+2.3}$ & $0.6_{-0.2}^{+0.3}$ & $0.3_{-0.3}^{+2.6}$ & $1.2_{-0.5}^{+1.9}$ \\
            \noalign{\smallskip}
            $kT_\mathrm{hard}$ (keV) & $1.46_{-0.05}^{+0.17}$ & $2.2_{-0.2}^{+0.3}$ & $1.46_{-0.15}^{+0.16}$ & $0.9_{-0.2}^{+0.8}$\\
            \noalign{\smallskip}
            Norm. ($10^{-2}$) & $3.1_{-1.5}^{+0.5}$ & $0.22_{-0.09}^{+0.13}$ & $3.0_{-0.9}^{+1.4}$ & $3.6_{-3.3}^{+13.6}$\\
            \noalign{\smallskip}
            $\Gamma$ & -- & -- & 0.38 & 0.38 \\
            \noalign{\smallskip}
            $E_\mathrm{cut}$ (keV) & -- & -- & 4.3 & 4.3 \\
            \noalign{\smallskip}
            Norm. ($10^{-5}$) & -- & -- & 2.8\tablefootmark{a} & $6.8_{-2.6}^{+0.7}$\\
            \noalign{\smallskip}
            $F_\mathrm{tot}$ ($10^{-12}$ erg cm$^{-2}$ s $^{-1}$) & $3.06\pm0.03$ & $1.57\pm0.02$ & $3.06\pm0.03$ & $1.64\pm0.02$ \\
            \noalign{\smallskip}        $F_\mathrm{cutoff}/F_\mathrm{tot}$ & -- & -- & $0.013\tablefootmark{a}$ & $0.50\pm0.02$ \\
            \noalign{\smallskip}
            $L_X$ ($10^{39}$ erg s $^{-1}$) & $4.77\pm0.05$ & $2.44\pm0.03$ & $4.77\pm0.05$ & $2.56\pm0.03$ \\
            \noalign{\smallskip}
            $\chi^2$/dof & 1220.12/1153 & 884.20/832 & 1220.11/1152 & 880.08/831 \\
            \noalign{\smallskip}
            n.h.p.\tablefootmark{b} & 0.083 & 0.102 & 0.08 & 0.116 \\
            \hline\hline
         \end{tabular}
     $$ 
   \tablefoot{
   The unabsorbed X-ray fluxes and luminosities are in the 0.3--10 keV range, while the flux ratio $F_\mathrm{cutoff}/F_\mathrm{tot}$ is computed in the range 0.3--40 keV, for the purpose of comparison with \citet{Walton2018}. The cut-off power law parameters, except the normalisation, are frozen to the best-fit ones of the SLL regime obtained using the \nustar{} dataset; the interstellar absorption is fixed to $N_H=9.8\times10^{20}$ cm$^{-2}$ (G21). Errors are at 90\% confidence level.
   \tablefoottext{a}{Upper limit.}
   \tablefoottext{b}{The null hypothesis probability (n.h.p.) is the probability of the observed data being drawn from the model, for a certain value of the $\chi^2$ and the degrees of freedom (dof). In the present work, we chose the common threshold of 0.05 and, hence, a fit with n.h.p. $>$0.05 can be considered statistically acceptable.}
   }
\end{table*}

\section{Temporal analysis}
\label{sec:timing_analysis}

\begin{figure}
    \resizebox{\hsize}{!}
    {\includegraphics[]{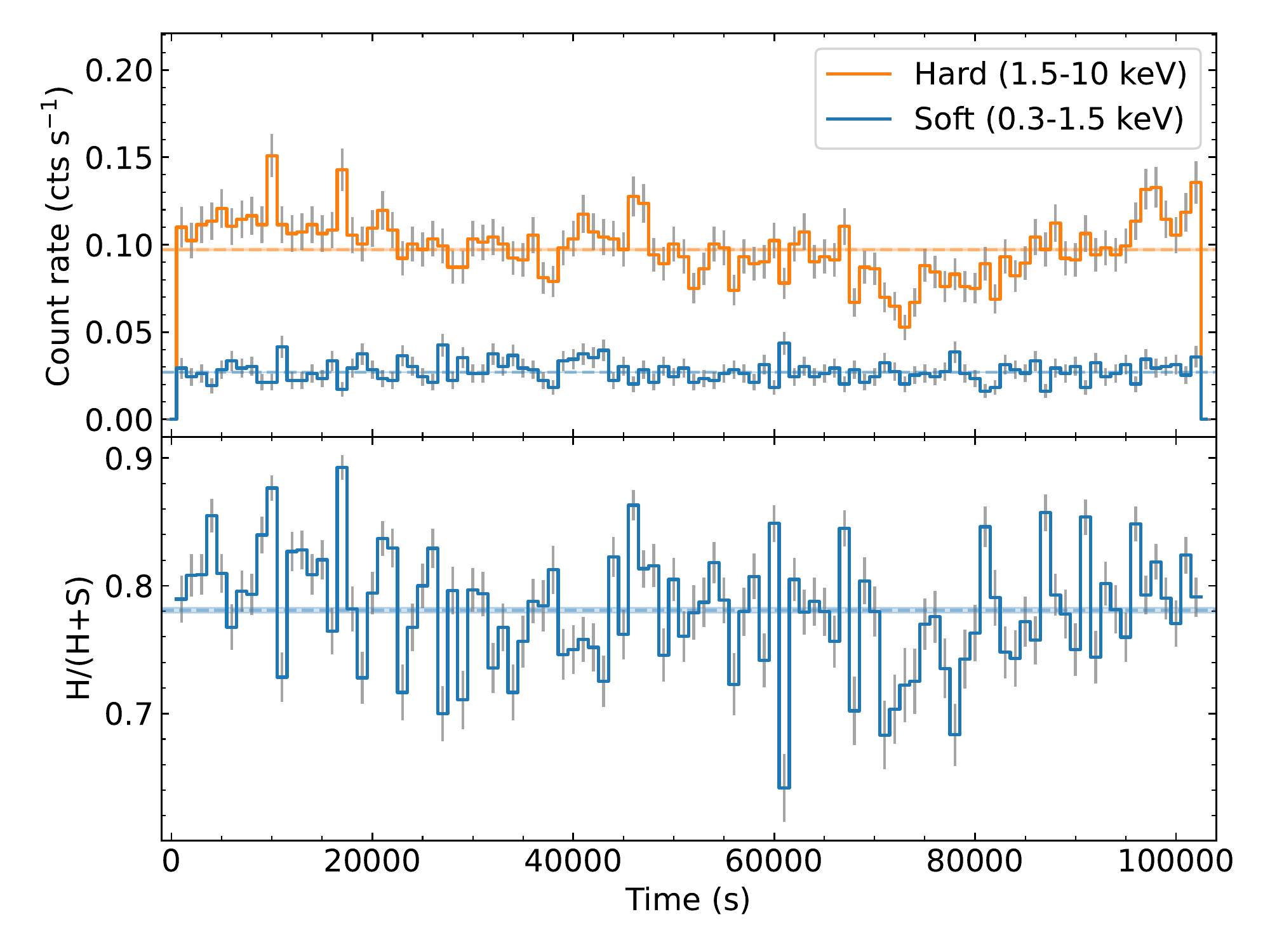}}
    \caption{Light curves and HR of \chandra\ ObsID 18047. \textit{Top panel}: Light curves in the hard (orange) and soft (blue) energy bands, binned at 1000 s, with dashed lines representing the mean count rate of each light curve. \textit{Bottom panel:} HR, computed as the number of counts in the hard band divided by the total number of counts. The dashed line corresponds to the mean HR.}
    \label{fig:18047hr}
\end{figure}

As for other ULXs, the X-ray light curves of M81 X-6 show short-term variability ranging from a few tens to thousands of seconds. Such variability is stronger in the hard band than in the soft one, as seen in many ULXs \citep[e.g.][]{Middleton2011,Sutton2013, Weng&Feng2018}. An example is shown in Fig.\,\ref{fig:18047hr}, which  also displays the HR as the fraction of counts in the hard band over the total counts in the broad energy band, for an easier interpretation. The same trend was also observed in \swift/XRT data by \cite{Weng&Feng2018} (cf. Sect. \ref{sec:discussion}) and it is consistent with the lack of variability of the soft spectral component presented above.

To quantify the overall degree of variability of the source, we computed the fractional variability $F_\mathrm{var}$ (i.e. the square root of the normalised excess variance) as in \cite{Vaughan2003} for the \chandra\ data, dividing the light curves into segments of 20 ks, corresponding to the exposure of the shortest observation, and averaging the obtained values. Due to the good time intervals and the intrinsic quality of the data, only five datasets allowed for a correct estimation of the fractional variability (see Table\,\ref{tab:fvar}). All five values are consistent with one another within the errors, so no trend of $F_\mathrm{var}$ with hardness or luminosity can be established.

As stated in Sect. \ref{sub:xmmdata}, \xmm\ observations are too short for a meaningful timing analysis. However, we did compute the fractional variability for the old \xmm\ ObsID. 0111800101, taken in 2001 (see G21), which has a good exposure of 73.6 ks for the pn. It resulted in $F_\mathrm{var}=0.037\pm0.005$, computed over four 20-ks intervals, and hence consistent with the values of the \chandra\ observations.

Unfortunately, with the present data, we could not conduct a search for pulsation of the order of seconds, as found for PULXs, since none of the \xmm/pn observations has a sufficient number of counts \citep[see][]{RodriguezCastillo2020}. Nonetheless, we determined the upper limit on the pulsed fraction (PF) for the \xmm/pn observation of 2001. We used the software {HENDRICS}, version 7.0.2 \citep{Bachetti2018} on the barycentred events at the \chandra\ position of M81 X-6 \citep{Swartz+2003} on the 0.3--12 keV band and searched for purely sinusoidal profiles with the $Z^2_n$ statistic (i.e. $n$=1), using 12 bins for each folding frequency and scanning the 0.1--8 Hz frequency range (similar to other ULXs). We used the option \textit{fast}, which automatically optimises the $f$--$\dot{f}$ search space. We did not detect any signal in any of the long $\sim$20 ks good time intervals (GTIs), and placed an upper limit on the PF $\sim$10.5\% at a 90\% confidence level. Combining three of the long continuous GTIs yielded a more stringent upper limit of $\sim$7.2\%. We did not compute PF upper limits for the other \xmm/pn observations, due to their short exposures.

\begin{table}
      \caption[]{Fractional variability of some of the \chandra\ observations. \xmm\ ObsID 01118001001 was taken in 2001, and it is analysed in G21.}
         \label{tab:fvar}
     $$ 
         \begin{tabular}{cccc}
            \hline\hline\noalign{\smallskip}
            ObsID & $F_\mathrm{var}$ & n. segments & State \\
            \noalign{\smallskip}
            \hline
            \noalign{\smallskip}
            18047 & 0.09$\pm$0.02 & 3 & HHL \\
            18817 & 0.04$\pm$0.06 & 2 & SLL \\
            18875 & 0.07$\pm$0.03 & 1 &  HHL \\
            19685 & 0.04$\pm$0.07 & 1 & SLL \\
            19982 & 0.04$\pm$0.04 & 1 & HHL \\
            \noalign{\smallskip}
           \hline
            \noalign{\smallskip}
            0111800101 & 0.037$\pm$0.005 & 4 & HHL \\
            \noalign{\smallskip}
            \hline\hline
         \end{tabular}
     $$ 
\end{table}

\begin{figure}
    \resizebox{\hsize}{!}
    {\includegraphics[]{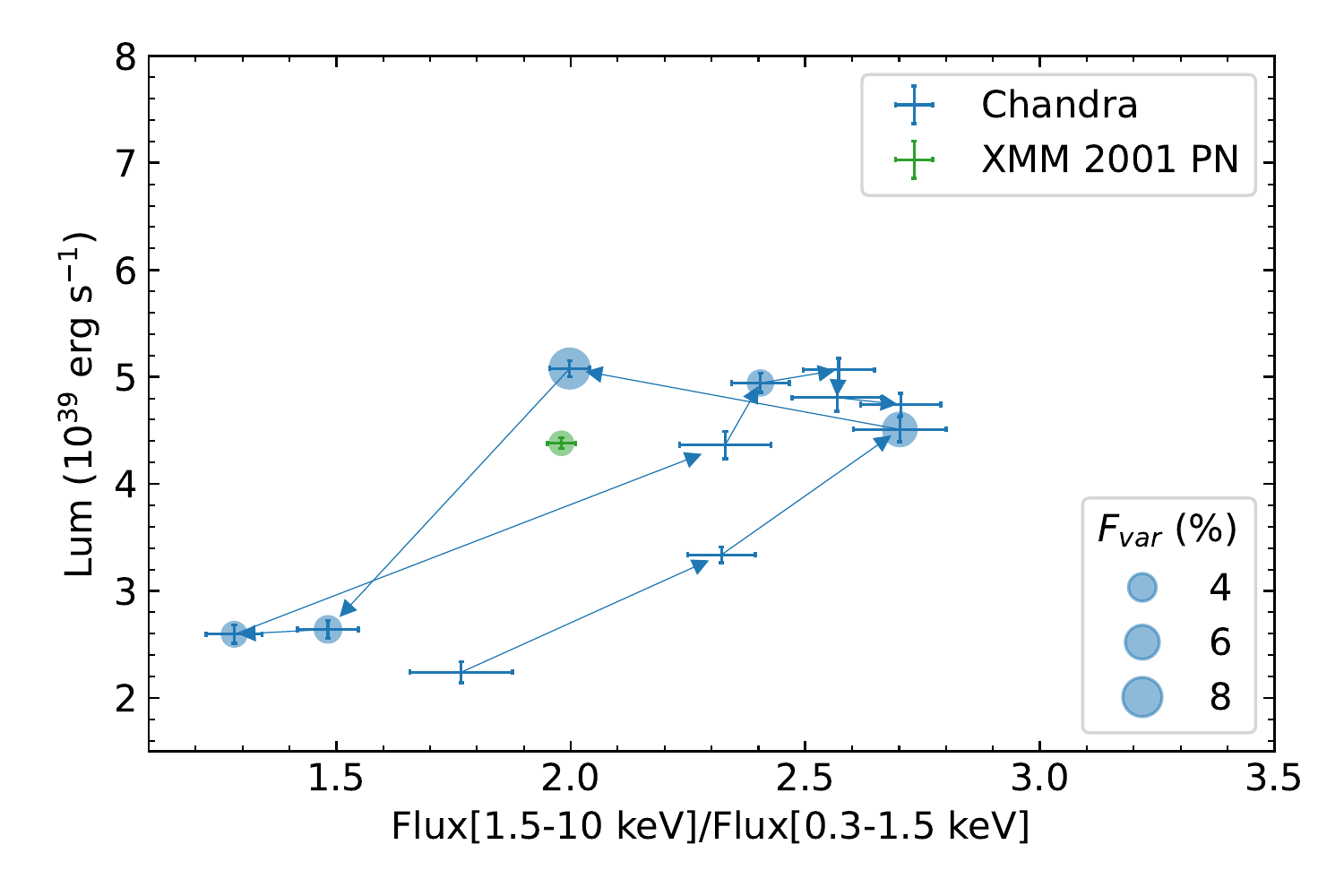}}
    \caption{HID of the  \chandra\ observations and one \xmm\ observation taken in 2001 (G21). The circle size is proportional to the fractional variability, only shown on observations for which it could be constrained.}
    \label{fig:fvar}
\end{figure}

\section{Discussion}
\label{sec:discussion}

We have presented a thorough analysis of all available X-ray data of the ULX M81 X-6 taken with \chandra, \xmm, \swift/XRT, and \nustar\ in the last decade. Older \xmm{} archival observations, up to 2012, were already analysed by G21. Here, we built on their work and presented an up-to-date analysis of the source. 

\subsection{M81 X-6 spectral states}

According to the general classification of ULXs \citep{Sutton2013}, with its HR always $\gtrsim$1 (see Figs.\,\ref{fig:LumFluxXMM} and \ref{fig:Swifthardness}), M81 X-6 would fall into the general category of HUL sources. However,  as already noted in G21, within its HUL state it oscillates between two main regimes, which we defined as HHL (HR$\sim$2--3 and $L_\mathrm{X}\sim4-6\times10^{39}$ \ergs{}) and SLL states (HR$\lesssim$2 and $L_\mathrm{X}\sim3\times10^{39}$ \ergs{}; see Fig.\,\ref{fig:LumFluxXMM} and Sect. \ref{sec:spectral_analysis}). A comparison with the much more extended \swift/XRT data sample (Fig.\,\ref{fig:Swifthardness}) confirms that the source oscillates over the years between these two main states, without any hint of any further state, as opposed to, for example, Holmberg II X–1 and NGC 5204 X–1, \citep{Gurpide2021b}. The two different types of track found for the ULXs in the sample of G21 led the authors to suggest that the nature of the accretor (BH vs NS) may be responsible for the spectral differences. As a matter of fact, the track on the HID of M81 X-6 is similar to those of other ULXs (e.g. M83 ULX1, Circinus ULX5, Holmberg IX X-1, NGC 55 ULX1, and M51 ULX8; see G21) and remarkably similar to the PULX NGC 1313 X-2 track \citep[see also][]{Sathyaprakash2022}. It also shares the same range of hardness (HR$\sim$1--3) and luminosity ($L_X\sim$2--10$\times10^{39}$ \ergs) with NGC 1313 X-2, leading to the idea that M81 X-6 could host a NS as well. Under the hypothesis of a NS central engine, a lack of detected pulsation can be attributed to the relative inclination of the rotational axis of the system and the direction of the line of sight (G21), to the scattering within the wind cone \citep[e.g.][]{Mushtukov2021}, or to the relative position of the magnetospheric and the spherisation radii \citep[e.g.][]{Walton2018}. A separate analysis of the two states would hence shed light on the geometry of the system and on the physical processes at play, as we discuss below and in the following sections.

We sorted all the \chandra\ and \xmm\ observations by state and fitted the two groups, to better constrain the parameters of the model. The simple model of two thermal discs that we tried at first resulted in a scenario in which the temperature of the hard component was higher in the SLL state than in the HHL state. Motivated by this possible contradiction and by the presence of small residuals at energies $\gtrsim8$ keV, we decided to add a further component to the fits, to better model the harder part of the spectra. To this purpose, we chose a cut-off power law, which we constrained thanks to the \nustar\ observation, taken when the source was in the SLL regime. With this new model, we found a temperature of the soft disc component equal to $\sim$0.4 keV in both states, while the temperatures of the hard disc component are $\sim$1.5 keV and $\sim$1 keV, in the HHL and SLL states, respectively. We must stress once again that the lack of any \nustar{} observation of M81 X-6 in the HHL at high energies prevented us from correctly modelling the cut-off power law in this state. Nevertheless, we resolved to use this component to fit the HHL spectra as well, under the assumption that the disc temperatures are more meaningful.

\subsection{M81 X-6 as a NS-ULX}

\cite{Walton2018b} found that the spectra of the PULXs NGC 5907 ULX and NGC 7793 P13 show a hard excess above $\sim$10 keV when fitted with two thermal disc models. Based on phase-resolved spectroscopy, the authors determined that this hard excess was attributed to the pulsed component, modelled with a cut-off power law. Moreover, \cite{Walton2018} found that those PULXs had a larger contribution to the 0.3--40 keV flux from this component (typically above 50\%), compared to ULXs for which pulsations had not been found. In light of that, we computed the ratio between the flux of the cut-off power law and the total flux $F_\mathrm{cutoff}/F_\mathrm{tot}$ (Table \ref{tab:BestFitGroups}), using the same energy range of \cite{Walton2018}. We found a ratio of 50\% in the SLL regime, where we had \nustar\ coverage. It is interesting to note that the value found for the SLL sits in between those seen in PULXs and the rest of the ULX population with broadband spectroscopy available. Given the similarities of M81 X-6 with NGC 1313 X-2 and the low PF of the latter (the lowest in the PULX population), our analysis would support that M81 X-6 indeed hosts a NS with a small PF. We also found that the $F_\mathrm{cutoff}/F_\mathrm{tot}$ ratio drops to below 2\% in the HHL regime. This result is more uncertain, given that we had to assume the same parameters for the cut-off power law as for the SLL state, due to the lack of high-energy ($>$10 keV) coverage in this state. Nonetheless, it suggests that pulsations should be more readily observable in the SLL state, similarly to NGC 1313 X-2, where pulsations were detected in the low-luminosity state \citep[][G21]{Sathyaprakash2019}. Moreover, the PF for the \xmm{} observation taken in 2001 resulted in an upper limit of $\lesssim$10\% (cf. Sect. \ref{sec:timing_analysis}), consistent with the PF of other ULXs \citep{Fuerst2016, RodriguezCastillo2020}, but higher, for instance, than the PF of NGC 1313 X-2 \citep{Sathyaprakash2019}. Whether this low PF is specific to this spectral state or applies to all states remains to be seen.

\subsection{Long-term variability}

M81 X-6 shows a small variability of the broad-band light curves throughout its evolution, with low values of fractional variability ($F_\mathrm{var}\lesssim10$\%; see Table \ref{tab:fvar}), close to those of NGC 1313 X-2 \citep{Robba2021}, for which a correlation of the fractional variability with the luminosity has been observed \citep[it has to be noted, however, that other ULXs show higher values; see e.g.][]{Middleton2015,Salvaggio2022}. Unfortunately, in our case the quality of the data and/or the short length of the observations, especially those made with \chandra{}, meant large error bars for the fractional variability so that it is difficult to infer a trend within the two states. Nonetheless, one of the oldest \xmm/pn observations (ObsID. 0111800101; G21) was long enough to constrain its fractional variability with a sufficiently high precision. This particular observation took place when the source was in a transitional state close to the HHL regime, but its low fractional variability of 0.037$\pm$0.005 alone would seem to contradict the trend found for NGC 1313 X-2. More observations would be of interest to confirm or rule out this trend.

Nevertheless, M81 X-6 exhibits a bi-modal distribution on the HID, already noticed by \cite{Weng&Feng2018} for \swift/XRT data up to August 2017 (cf. Fig.\,\ref{fig:Swiftlc}, right panel), with a strong quasi-periodic modulation in the hard band (1--10 keV), but not in the soft one (0.3--1 keV). These spectral changes seem to be associated with a quasi-periodicity of $\sim$110 days \citep{Lin2015}. \cite{Lin2015} also reports another periodicity at $\sim$ 370 days, but we note it is uncertain as it was only supported by two cycles and, given the slow timescale, could be a spurious feature. This bi-modal behaviour, common to other investigated ULXs (e.g. Holmberg IX X-1 and X-2), has been attributed to precession of the accretion flow, responsible for the modulation of the hard X-rays coming from the innermost regions, while the isotropic soft emission arising from the outflows and the photosphere of the outer disc would remain more or less constant (cf. Sect. \ref{sub:precession}).

Alternatively, a bi-modal distribution, with variations of more than an order of magnitude in luminosity observed for other ULXs, has been attributed to the propeller effect \citep{Earnshaw2018,Song2020}. However, the difference in luminosity of M81 X-6 between the HHL and the SLL regimes is slightly less than a factor of two. The ratio of the luminosities in/out of the propeller regime can be estimated as $\simeq170~P^{2/3}~M_{1.4}^{1/3}~R_6^{-1}$ \citep[][Eq.\,5]{Tsygankov2016}\footnote{Note that this formulation assumes sub-Eddington accretion and might be erroneous for ULXs. A correction of this formula is beyond the goals of this paper \citep[see][]{Middleton2022}.}, where $P$ is the NS rotational period, $M_{1.4}$ its mass in units of 1.4$M_\odot$, and $R_6$ it radius in units of 10 km. Using the canonical values for a NS of 1.4$M_\odot$ and 10-km radius, a quick estimation would lead to $P\sim1.28$ ms, far too small compared to the spin period of known PULXs. Hence, assuming that all NS-ULXs share the same properties as those already observed, we can exclude propeller transitions as the mechanism responsible for the variability. This conclusion is obviously based on a rather small sample of PULXs and on the limitations of current X-ray instruments in detecting those with shorter periods. Increasing the sample of PULXs will be crucial to properly investigate the role of the propeller regime in ULXs.

\subsection{The nature of the super-orbital period}
\label{sub:precession}

If we consider a scenario in which the soft thermal component represents the outflow photosphere, where the temperature is a function of the mass-accretion rate \citep{Poutanen2007}, then the lack of variation in $T_\mathrm{soft}$ between the two regimes suggests that the luminosity and spectral variations of the source are not driven by changes in the mass-transfer rates, in line with G21. The bi-modal variability and the presence of a variable $\sim$110 days quasi-periodicity \citep{Lin2015} may therefore imply that these changes are driven by precession of the accretion disc. The nature of this super-orbital period in ULXs is still uncertain \citep[e.g.][]{Vasilopoulos2020}. Here, we explore two possible scenarios: Lense-Thirring precession and the torque from the NS magnetic dipole. We did not explore free precession \citep[see e.g.][]{Vasilopoulos2020}, because the variability in the super-orbital periods of some sources \citep[e.g. M51 ULX7;][]{Brightman2022} would seem to disfavour this scenario.\\

Lense-Thirring precession is caused by the misalignment of the rotational axis of the system and NS spin axis, which induces frame-dragging and oscillations in the disc. \citet{Middleton2019} suggested that BH- and NS-ULXs should be distinguishable on the basis of the temperature from the spherisation radius ($R_\mathrm{sph}$) and super-orbital period, given the physical differences between these two objects (e.g. spin, mass, presence of a surface, etc). Additionally, it may be possible to offer some constraints on the magnetic field of the source ($B$) and its spin ($P_\mathrm{NS}$). To this end, under the assumption that Lense-Thirring precession drives the variability of the source, we employed the formulae presented by \citet{Middleton2019} \citep[see also e.g.][]{Poutanen2007,Mushtukov2019, Vasilopoulos2019, Middleton2018, Fragile2007}, setting all relevant unknown parameters (such as the accretion efficiency or the ratio of the asymptotic wind velocity relative to the Keplerian velocity, $\beta$) as in the original work of \cite{Middleton2019}, unless stated otherwise. We note that we assume the precession to extend until the photospheric radius \citep[as for][]{Middleton2019}, but this should be regarded as an upper limit for the radius at which the flow is able to precess as a solid body, as the bending wave might not be able to propagate if the outflow becomes supersonic at smaller radii \citep[see][]{Middleton2019}.

A complete modelling of the Lense-Thirring precession requires the tuning of two more parameters: the dimensionless NS spin, $a_*$, and the fraction of energy that powers the outflow, $\varepsilon_\mathrm{wind}=L_\mathrm{wind}/(L_\mathrm{wind} + L_\mathrm{rad})$, with $L_\mathrm{wind}$ the kinetic luminosity of the wind and $L_\mathrm{rad}$ the observed radiative luminosity \citep{Poutanen2007}. Based on the periods found in ULXs \citep[e.g.][]{Bachetti2014,Israel2017a, RodriguezCastillo2020}, $a_*$ should be $\ll$ 0.01, as already pointed out by \cite{Middleton2019}. Assuming canonical values for the moment of inertia of the NS ($I$ = 10$^{45}$ g cm$^2$) and its mass ($M$ = 1.4 $M_\odot$), we find that 3 $\times$ 10$^{-2}$ $>$ $a_*$ $>$ 3 $\times$ 10$^{-5}$ should approximately cover the range of observed periods in ULXs ($\sim$ 0.1 s $<$ $P$ $<$ 10 s). For $\varepsilon_\mathrm{wind}$, we considered values of 0.2 and 0.8 to cover the range of values derived from numerical simulations \citep[e.g.][]{Jiang2014} and some observational constraints derived from shock-ionised optical bubbles around ULXs, as for, for example, NGC 1313 X-2 \citep{Pakull2002}, IC~342~X-1 \citep{Cseh2012}, NGC 5585 X-1 \citep{Soria2021}, and NGC~1313~X-1 \citep{Gurpide2022}.

\begin{figure*}
    \centering
    \includegraphics[width=0.48\textwidth]{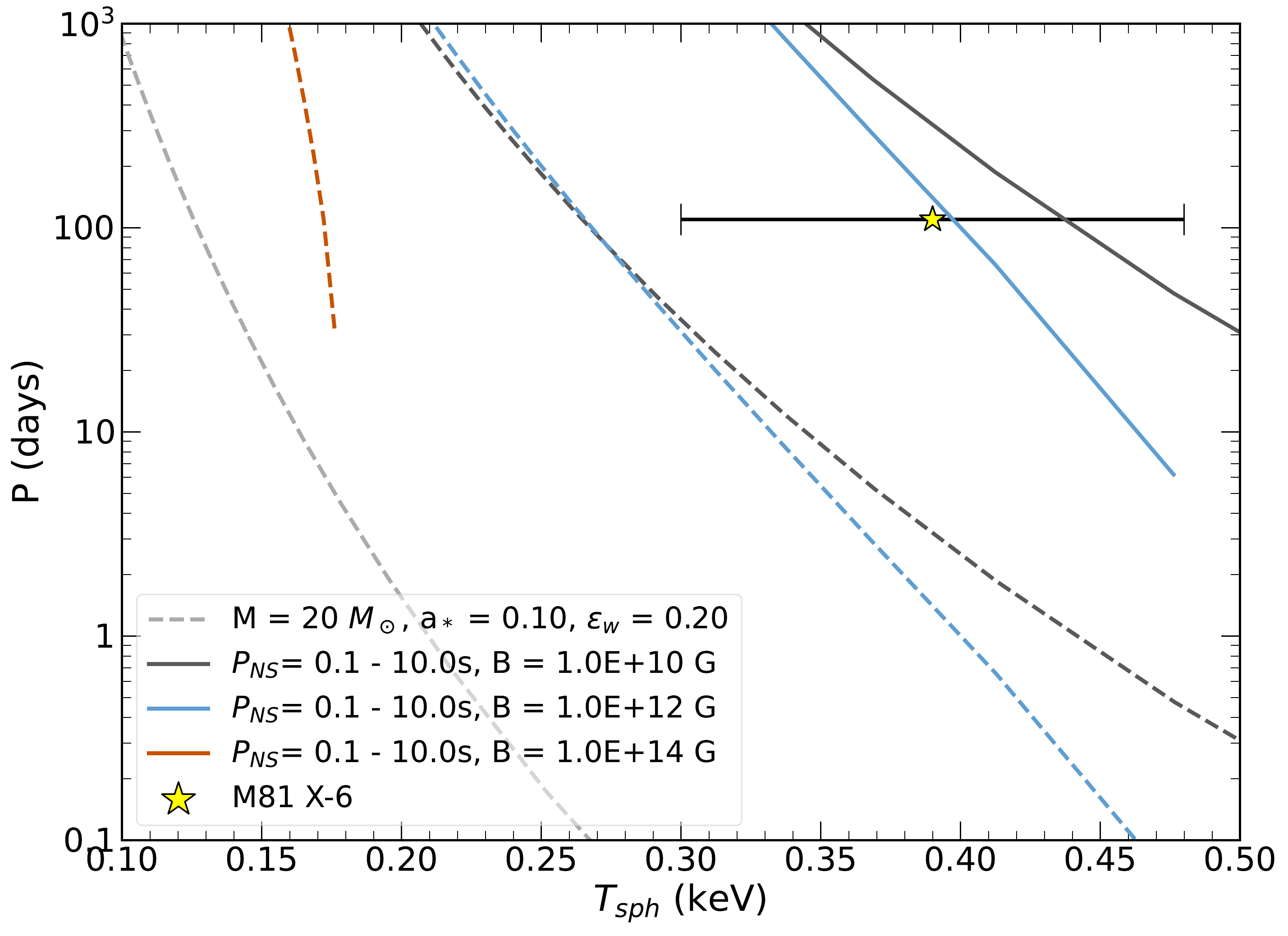}
    \includegraphics[width=0.48\textwidth]{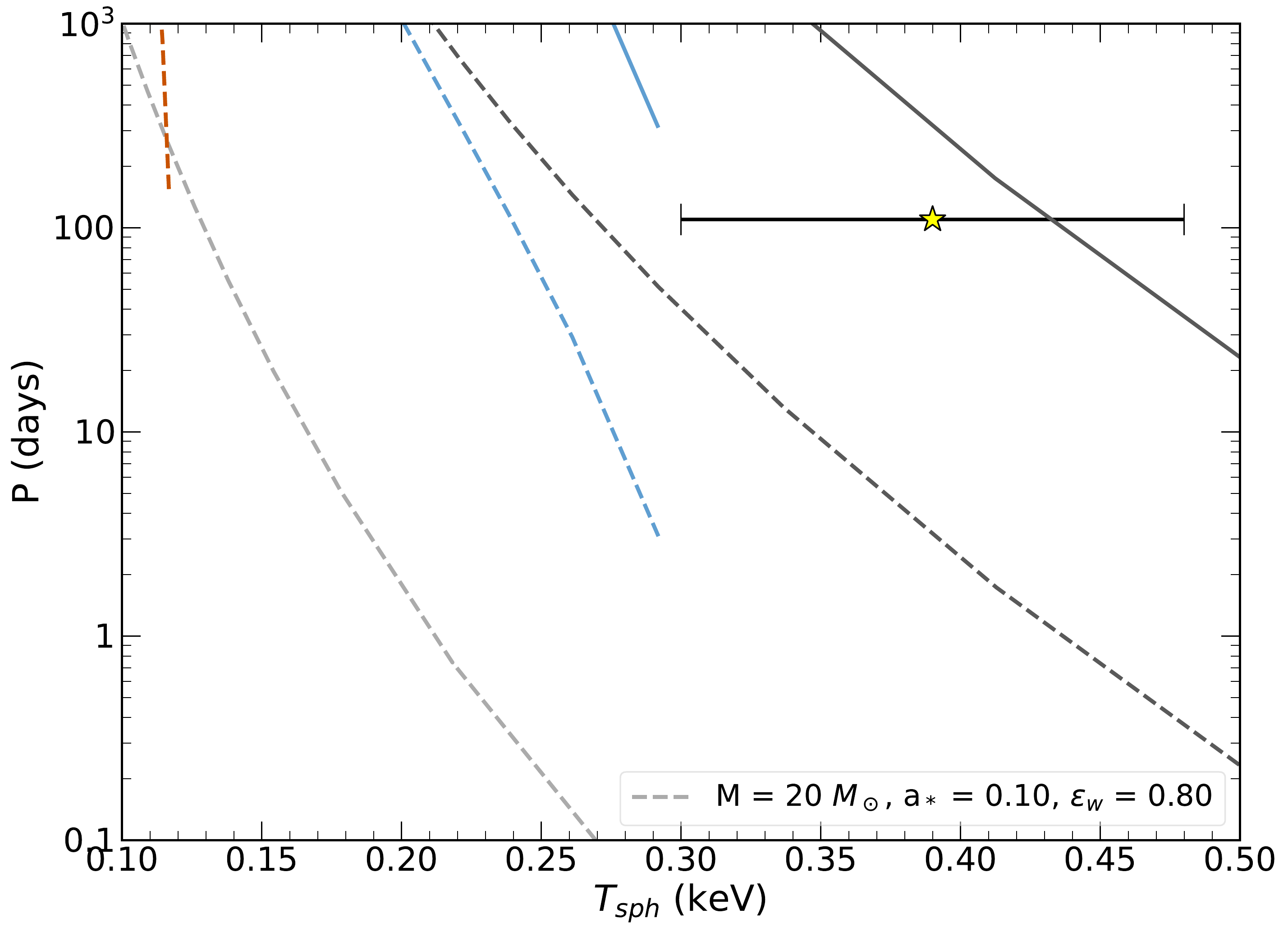}
    \caption{Lense-Thirring accretion plane adapted from \citet{Middleton2019}. The left and right panels show the resulting $T_\mathrm{sph}$ and precessing period of the wind for $\varepsilon_\mathrm{wind}$ = 0.2 and 0.8, respectively. The values expected for a 20\,$M_\odot$ BH with a dimensionless spin of 0.1 are shown in dashed dark blue and could be considered as a rough upper limit for the location of BHs in the plane. The rest of the lines show the values for NSs with different dipole magnetic field strengths, as indicated. The solid and dashed lines show the effect of varying the spin of the NS from 3 $\times$ 10$^{-5}$ to 3 $\times$ 10$^{-3}$, which we have converted to periods assuming canonical values for the moment of inertia and mass of the NS (see the text for details) and represent typical values observed in ULXs. }
    \label{fig:ls_plane}
\end{figure*}

The resulting Lense-Thirring plane is shown in Fig.\,\ref{fig:ls_plane}, together with the position of M81 X-6 in it and the expected values for a 20$M_\odot$ BH with a moderate spin of 0.1 for reference. We note that attributing the temperature derived from our simple modelling to the temperature from $R_\mathrm{sph}$ is certainly problematic as (a) our modelling is not able to capture the complexity of the accretion flow and (b) it is unlikely that the hottest temperature returned by the \texttt{diskbb} from a range of black bodies coincides with the temperature from $R_\mathrm{sph}$. Because of the latter reason, our naive expectation would be that our estimate for $T_\mathrm{sph}$ constitutes an upper limit. These caveats should be borne in mind when interpreting these results and therefore our estimates should be consider semi-quantitative. Improvement in theoretical modelling of PULXs should help reduce these uncertainties in the future.

From Fig.\,\ref{fig:ls_plane}, we see that the position of M81 X-6 on the plane indicates a weakly magnetised NS ($B < 10^{10}$ G) as the accretor. We note that lower magnetic field values would yield equivalent results for M81 X-6 as those found for $B=10^{10}$ G. This is because for low magnetic field values ($B < 10^{10}$ G) and accretion rates with $T_\mathrm{sph} \lesssim 0.6$ keV we find already that $R_\mathrm{mag} \xrightarrow{} R_\mathrm{ISCO}$ (i.e. the magnetospheric radius $R_\mathrm{mag}$ is pushed all the way to the innermost stable orbit $R_\mathrm{ISCO}$) and therefore the solutions become independent of the magnetic field. This implies that our magnetic field estimate constitutes an upper limit, which could be as high as 10$^{12}$ G if we consider the curves for $\varepsilon_w$ = 0.2. We note however that mostly $B <$\,$10^{10}$\,G would seem favoured as $\varepsilon_w$ = 0.2 constitutes a limiting case and high values would rule out the case for $B$ $\sim$ 10$^{12}$ G. Moreover, since higher values of $\varepsilon_w$ have the effect of increasing the size of the magnetosphere, as more mass is lost into the outflow, for $\varepsilon_w$ $>$ 0.3, a magnetic field with $B = 10^{12}$ G in M81 X-6 would truncate the disc before becoming supercritical (this effect can be seen on the Fig.\,\ref{fig:ls_plane} as the regions where the lines cut, in particular on the right panel), which would be inconsistent with this scenario. 

Therefore, according to this scenario and considering the super-orbital period $P=$ 111 d \citep{Lin2015}, $T_\mathrm{sph}\sim$0.4 keV, $\varepsilon_\mathrm{wind}=0.2$ and using $B=10^{10}$ G, we would expect for M81 X-6 a spin period $P_\mathrm{NS}\sim$4.5 s, a mass-accretion rate at the magnetospheric radius $\dot{m}_0$ = 42, a mass-accretion rate at the inner disc radius $\dot{m}_\mathrm{in} \sim 37$, a magnetospheric radius $R_\mathrm{mag} \sim 13R_\mathrm{g}$, a spherisation radius $R_\mathrm{sph} \sim 302 R_\mathrm{g}$, and an outer atmosphere radius $R_\mathrm{out} \sim 758R_\mathrm{g}$, where $R_\mathrm{g}$ is the gravitational radius ($R_\mathrm{g} = GM_{NS}/c^2$). For $\varepsilon_\mathrm{wind} = 0.8$, we found $P_\mathrm{NS}$ $\sim 4.7$s, $\dot{m}_0$ = 21, $\dot{m}_\mathrm{in} \sim 12$, $R_\mathrm{mag} \sim 18R_\mathrm{g}$, $R_\mathrm{sph} \sim 129 R_\mathrm{g}$, and $R_\mathrm{out} \sim 726 R_\mathrm{g}$. The first thing to note is that the predicted spin periods for the NS are within the range of those observed in ULXs and therefore not implausible. The values of $\dot{m}_0$ and $B$ found for both $\varepsilon_w$ seem to be within the allowed parameter space for the Lense-Thirring torque to dominate over the tidal torque according to \citet[][their Fig. 3]{Middleton2019}. Therefore, it might be reasonable to suggest that this torque might be at play in this source, and even be responsible for the precession. Finally, this scenario would support the identification of M81 X-6 as a weakly magnetised NS, where $R_\mathrm{sph} > R_\mathrm{mag}$, as suggested by G21. The overall low $R_\mathrm{mag}$, close to the $R_\mathrm{ISCO}$ = 6$R_\mathrm{g}$, coupled with the large $R_\mathrm{sph}$ might explain the lack of detectability of pulsations, due to a low PF component as a result of the small extent of the magnetised region or to the signal being scattered and lost in an accretion curtain \citep{Mushtukov2017}.\\

\begin{figure*}
    \centering
    \includegraphics[width=0.48\textwidth]{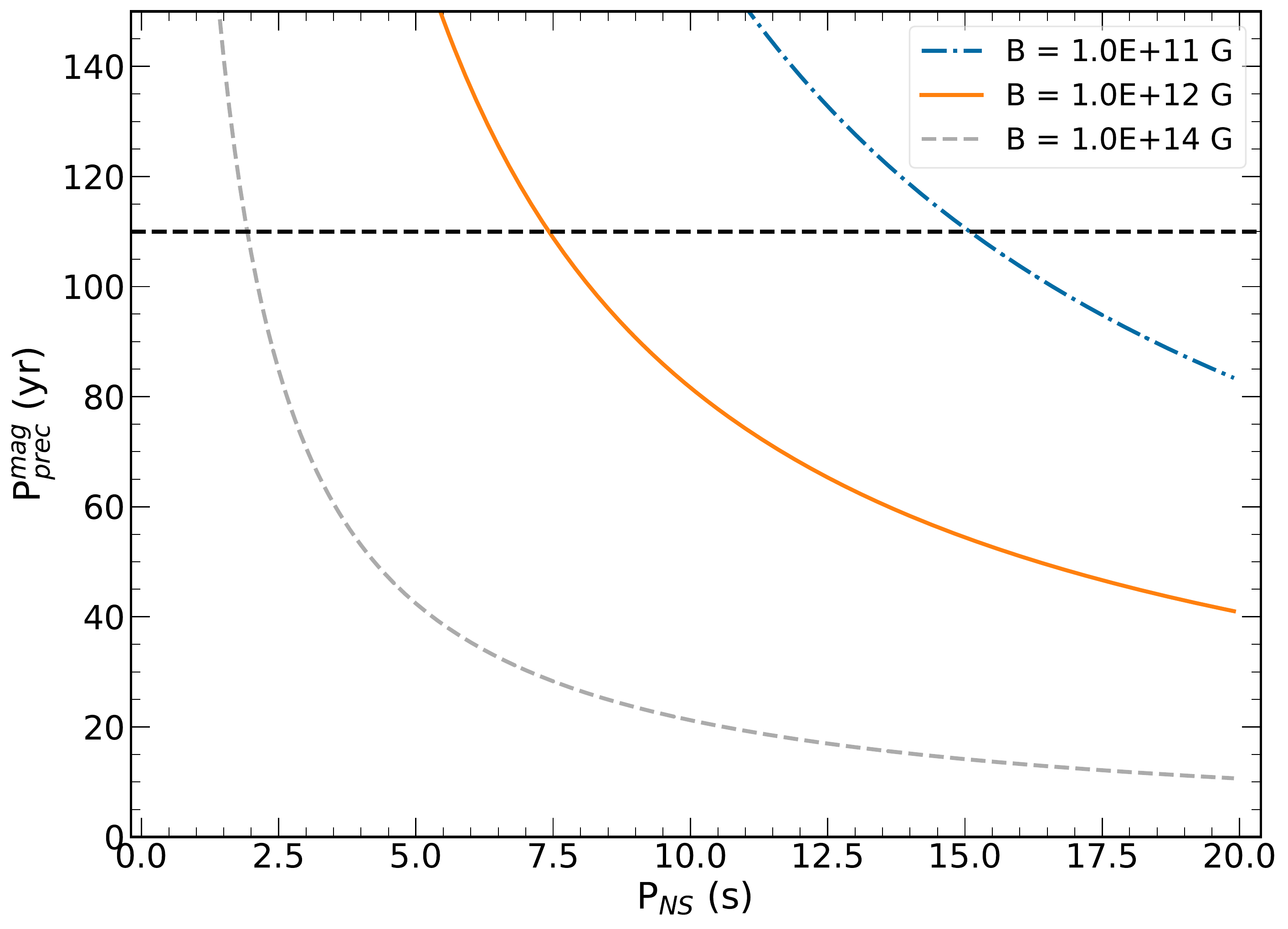}
    \includegraphics[width=0.48\textwidth]{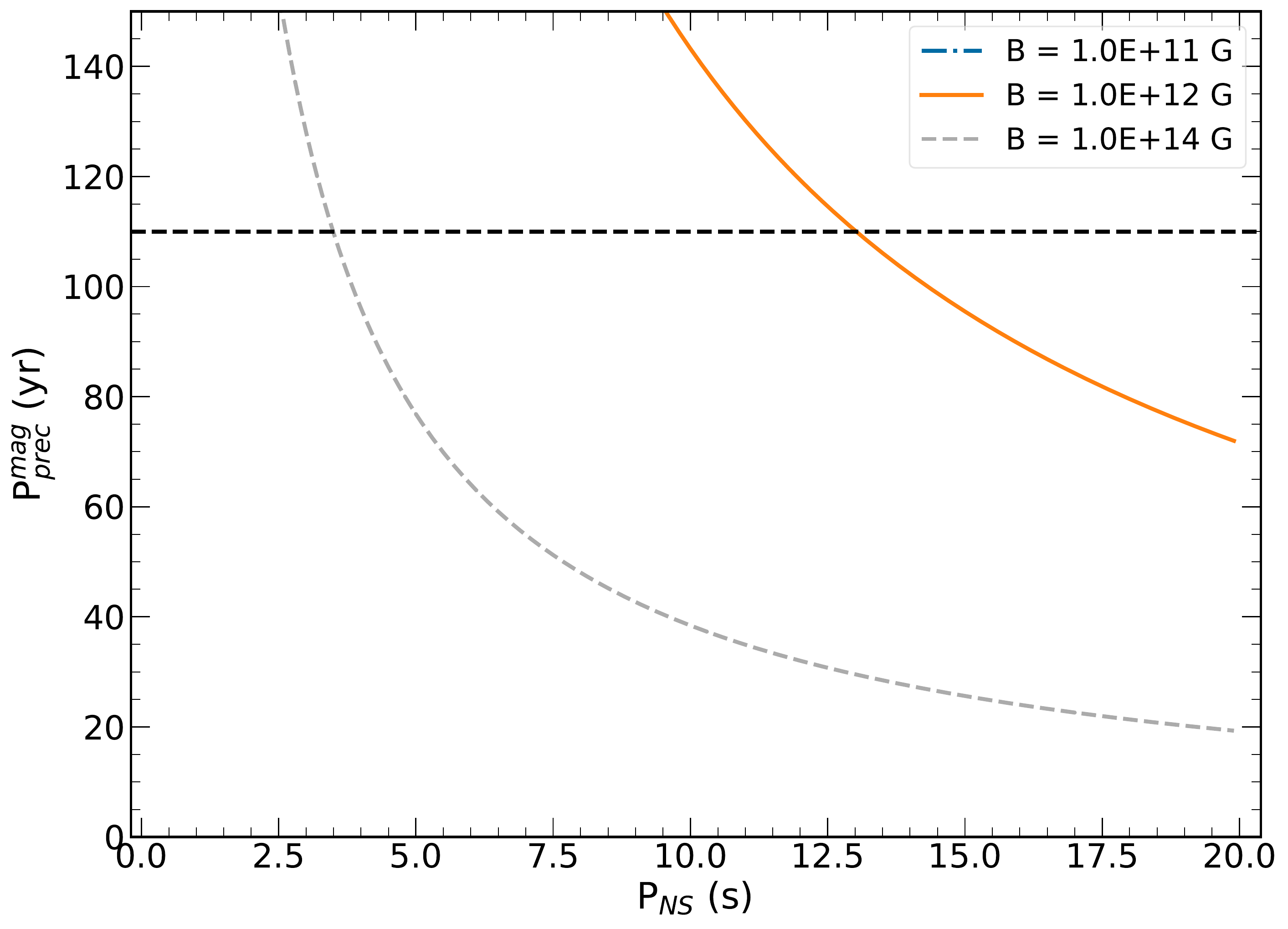}
    \caption{Precessing period due to the interaction of the NS magnetic field with the accretion disc \citep{Lipunov1980} as a function of the NS spin period calculated for different values of the NS dipole magnetic field. The left and right panels shows calculations for $\epsilon_w = 0.2$ and $\epsilon_w = 0.8$, respectively. The horizontal dashed line indicates the super-orbital period of 110 days of M81 X-6 \citep{Lin2015}, and we have considered the temperature of the soft component in the calculations of $R_\mathrm{mag}$. }
    \label{fig:prec_mag}
\end{figure*}

Given the relatively slow $P_\mathrm{NS}$ required for this scenario, however, it is possible that the magnetic torque due to the interaction from the NS magnetic field with the accretion disc \citep{Lipunov1980} will dominate over the Lense-Thirring torque, notably for magnetic fields $\gtrsim$10$^{11}$ G \citep[see][their Fig.\,8]{Middleton2018}. The exact calculation is not straightforward as the magnetic torque has only been estimated for a standard thin disc \citep[see][and references within]{Middleton2018}, but it is informative to consider whether this scenario could account for the super-orbital period of the source, as also explored by \citet{Vasilopoulos2020} in the case of the PULX M51 ULX7 \citep[see also][]{Mushtukov2017}. The precessing timescale of the NS can be estimated as \citep{Lipunov1980}
\begin{equation}
\begin{split}
    P^\mathrm{mag}_\mathrm{prec} = 1.5 &\times 10^4 \left (\frac{B}{10^{12}\,\mathrm{G}}\right)^{-2} \left (\frac{R_\mathrm{NS}}{10^6 \,\mathrm{cm}}\right)^{-2} \left (\frac{R_\mathrm{mag}}{10^8 \,\mathrm{cm}}\right)^3  \\
    \left (\frac{P_\mathrm{NS}}{1\,\mathrm{s}}\right)^{-1} & \left (\frac{I}{10^{45} \,\mathrm{gr}\,  \mathrm{cm}^2}\right) \frac{1}{\cos\psi(3 \cos^2\xi -1)}\, \mathrm{yr}
\end{split}
,\end{equation}
where $\psi$ is the angle between the NS rotational axis and the normal to the accretion disc and $\xi$ is the angle between the NS magnetic field and its rotational axis. $R_\mathrm{NS}$ is the radius of the NS at its surface, assumed to be 10 km. As above, we solve numerically for $R_\mathrm{mag}$, based on $\dot{m_0}$ estimated from the temperature of the soft component and for different NS magnetic fields. We set $\xi$ = $\psi$ = 30$^\circ$ for reference and carry out the calculations for $\epsilon_w$ = 0.2 and 0.8 as above and for $P_\mathrm{NS}$ = 0.1--20 s. We present the calculations in Fig.\,\ref{fig:prec_mag}, which can also be useful for future reference.

As opposed to the Lense-Thirring scenario, this model would require a strongly magnetised ($10^{12}$ G $< B < 10^{14}$ G) NS for M81 X-6, whereas the necessary spin period would be in line with those estimated under the assumption of Lense-Thirring precession. We thus see that estimating the magnetic field of the putative NS could be useful in putting constraints on the mechanism responsible for the super-orbital period. 

It must be noted that if this mechanism operates in the whole PULX population, it would require magnetic fields well above 10$^{15}$ G for those more rapidly spinning PULXs and/or with shorter super-orbital periods \citep[e.g. M51 ULX7; as already pointed out by][]{Vasilopoulos2020}. Therefore, if the same mechanism is not responsible for the super-orbital periods seen in other PULXs \citep[e.g.][]{Walton2016}, then we may be able to rule out this scenario for M81 X-6 too. We conclude that, from the mechanisms explored here, Lense-Thirring precession can explain the super-orbital period of M81 X-6, provided that the NS has a magnetic field $\lesssim$10$^{10}$ G.\\

\begin{figure}
    \resizebox{\hsize}{!}
    {\includegraphics[]{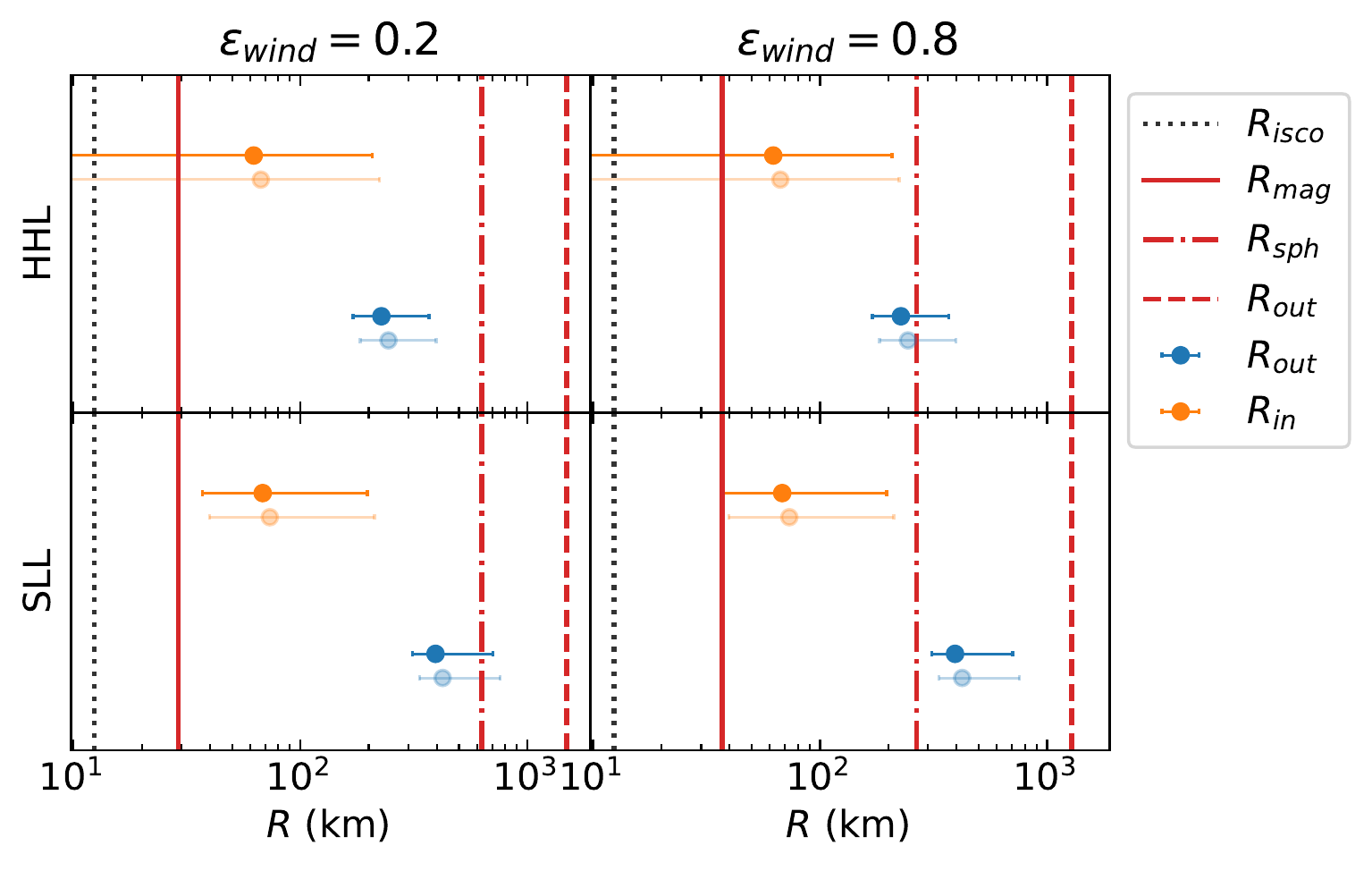}}
    \caption{Comparison of the physical sizes of the system derived from the best-fit multi-temperature disc components (dots) and from the Lense-Thirring precession model (vertical lines; see Sect.\,\ref{sub:precession} for details). The solid and transparent dots correspond to inclination angles of the system of 0 and 30 degrees, respectively. }
    \label{fig:radii}
\end{figure}

Finally, to come full circle, the characteristic radii derived above can be compared with the `apparent' inner emitting radii \citep{Kubota1998} of the two multi-temperature disc components. These can be computed from the normalisations of the two black bodies, for which we assume two extreme inclination angles of 0 and 30 degrees inferred in G21, as the source is viewed at low inclinations (see also G21). Results are shown in Fig.\,\ref{fig:radii}. In our interpretation of the system, the soft emission is arising from the outer disc and/or the outflows, between $R_\mathrm{sph}$ and $R_\mathrm{out}$, whereas the harder emission would come from the inner region from within $R_\mathrm{sph}$ down to $R_\mathrm{ISCO}$ for a BH-ULX or to $R_\mathrm{mag}$ for a NS-ULX. The temperature of the soft disc component, which corresponds to the temperature of the disc at the spherisation radius, has been directly employed in the Lense-Thirring model. Then, it is not surprising that the apparent inner radii obtained from the soft component are consistent with $R_\mathrm{sph}$ or fall within $R_\mathrm{sph}$ and $R_\mathrm{out}$. The only exception may be the case in the HHL state with $\varepsilon_\mathrm{wind}=0.2$ (Fig.\,\ref{fig:radii}, top left panel), where the apparent inner radius falls slightly before $R_\mathrm{sph}$, but, in any case, the soft-emission region would still develop in between the $R_\mathrm{sph}$ and the $R_\mathrm{out}$. On the other hand, we find relatively good agreement between $R_\mathrm{mag}$ and the radii derived from the hard emitting component, even though this component is not directly implied in the Lense-Thirring model. In all possible scenarios, the inner radii fall within $R_\mathrm{mag}$ and $R_\mathrm{sph}$. This tentative agreement may suggest that our simplistic spectral fitting model might be a good first order approximation and may strengthen our interpretation of the various emitting regions attributed to the spectral components. A more accurate comparison should take into account, for example, the fraction of energy deposited in the wind, the colour-temperature correction, the fact the photosphere is anistropic, etc., but this advanced modelling is far beyond the goal of this paper.

\section{Conclusions}

We have investigated the long-term spectral and temporal evolution of the ULX M81 X-6, finding that the source oscillates between two different regimes, characterised by different values of hardness and luminosity. A modelling of the spectra in the two regimes indicates a probable high-energy component, which can be attributed to the emission coming from the accretion column, under the hypothesis of a NS accretor. 

The two thermal components describing the broadband 0.3--10 keV spectra can be linked to the inner (higher temperature) and outer (lower temperature) regions of the accretion disc. The constant temperature of the softer of these components would suggest that the spectral changes are not due to variations in the mass-accretion rate. On the other hand, the amount of variation in luminosity is not high enough to justify transitions into or out of the propeller regime, assuming that M81 X-6 has the same properties of other known ULXs. Therefore, the transitions between the two spectral states can be attributed to a precession of the accretion disc. 

Assuming a NS accretor, we tested two possible scenarios: Lense-Thirring precession and precession due to the magnetic torque applied by the NS magnetic field onto the accretion disc. In the case of Lense-Thirring precession, we estimated the characteristic radii of the system, the magnetic field ($B\lesssim10^{10}$ G), and the expected spin period ($\sim$5 s) of the NS. In the case of a magnetic torque, a higher magnetic field ($B\sim10^{12}-10^{14}$ G) is required, with a similar spin period. However, the latter mechanism would be disfavoured because it would require unrealistically high values of the magnetic field strength ($B>10^{15}$ G) for known PULXs. 

Clearly, the direct detection of pulsations would be the strongest evidence in favour of the NS-ULX nature of M81 X-6. However, a direct detection is difficult, not only because of instrumental limitations (e.g. time resolution, sensitivity, exposure times, etc.), but also because of the intrinsic physics and geometry of ULXs. For instance, in our case, the inferred low fraction of the flux of the cut-off power law component over the total flux, coupled with the upper limit on the PF of $\sim$10\% found for one \xmm/pn observation, would suggest a scenario where the pulsation is somehow `diluted' by the wind cone in the outer photosphere. This may be a consequence of the small extent of the magnetosphere of the NS, compared to the size of the spherisation radius of the accretion disc, as inferred from the application of the Lense-Thirring precessing model (cf. Sect. \ref{sub:precession}). 

New observational data and numerical simulations may prove crucial to more broadly  understanding the properties of PULXs and thus discriminating between the two precession mechanisms. Moreover, future pointed longer observations of {M81 X-6} with \xmm{}, \nustar{}, or next-generation X-ray satellites, such as {XRISM} \citep{Xrism2020} or \textit{Athena} \citep{Athena2013}, are encouraged to systematically look for pulsations and strengthen the findings of this work.

\begin{acknowledgements}
This research has made use of data obtained from the \textit{Chandra} Data Archive and the \textit{Chandra} Source Catalogue, and software provided by the \textit{Chandra} X-ray Center (CXC) in the application package CIAO; of the \textit{NuSTAR} Data Analysis Software (NuSTARDAS) jointly developed by the ASI Space Science Data Center (SSDC, Italy) and the California Institute of Technology (Caltech, USA); of observations obtained with \textit{XMM-Newton}, an ESA science mission with instruments and contributions directly funded by ESA Member States and NASA. This research has also made use of the software \textit{HENDRICS}, based on the software package \textit{Stingray} \citep{Bachetti2022_stingray, Huppenkothen2019, Huppenkothen2019JOSS}. RA, NW, and OG acknowledge the support of the CNES for this work. The authors thank S.~Vaughan for clarifications on the computation of the fractional variability. 
\end{acknowledgements}

%
\bibliographystyle{aa} 
\bibliography{biblio} 
%

\end{document}